\documentclass[11pt]{article}
\usepackage{epsfig}
\usepackage{amscd}
\usepackage{amssymb}
\usepackage{amsfonts}
\usepackage{amsmath}
\usepackage{mathrsfs}
\usepackage{graphics,epsfig,psfrag}
 
\begin{document}

\makeatletter
\@addtoreset{equation}{section}
\makeatother
\renewcommand{\theequation}{\thesection.\arabic{equation}}
\renewcommand{\thefootnote}{\alph{footnote}}

\begin{titlepage}

\baselineskip =15.5pt
\pagestyle{plain}
\setcounter{page}{0}

\begin{flushright}
\end{flushright}

\vfil

\begin{center}
{\LARGE {\bf Chiral Anomalies}} \medskip\\
{\LARGE {\bf via}} \medskip \\ 
{\LARGE {\bf Classical and Quantum Functional Methods}}
\end{center}

\vfil

\begin{center}
{\large E. Gozzi$^*$, D. Mauro$^*$, A. Silvestri}\footnote{{\it Present Address}: Department of Physics, Syracuse University, \\
\vspace{-0.3cm}

\hspace{0cm} Syracuse, NY 13244, USA;
e-mail: {\it asilvest@physics.syr.edu }}\\
\vspace {1mm}
Dipartimento di Fisica Teorica, Universit\`a di Trieste, \\
Strada Costiera 11, Miramare-Grignano 34014, Trieste \\ $^*$and INFN, Trieste, Italy\\
e-mail: {\it gozzi@ts.infn.it} and {\it mauro@ts.infn.it} \\


\end{center}

\vfil

\begin{abstract}
\noindent In the {\it quantum} path integral formulation of a field theory model an anomaly arises when the functional
measure is not invariant under a symmetry transformation of the Lagrangian. 
In this paper, generalizing previous work done on the point particle, we show that even at the {\it classical} level we can give a path integral formulation for any field theory model. Since classical mechanics cannot be affected by anomalies, the measure of the {\it classical} path integral of a field theory must be invariant under the symmetry. The {\it classical} path integral measure contains the fields of the {\it quantum} one plus some extra auxiliary ones. So, at the classical level, there must be a sort of ``cancellation" of the quantum anomaly between the original fields and the auxiliary ones. In this paper we prove in detail how this occurs for the chiral anomaly. 
\end{abstract}

\begin{center}
{\it In memory of Bunji Sakita}
\end{center}
\end{titlepage}
\newpage

\section{Introduction}

The path integral formalism \cite{uno} for {\it Q}uantum {\it M}echanics (QM) and {\it Q}uantum {\it F}ield {\it T}heory (QFT) has turned out to be one of the most powerful tools in theoretical physics. It has been crucial not only in rewriting the old-fashioned perturbation theory via the ``modern" and easy to use Feynman diagrams, but also in the development of non-perturbative techniques like the ones based on the use of instantons, solitons etc. 

At the end of the Seventies it also helped in throwing new light on those important phenomena known as quantum anomalies \cite{due}. These symmetries are present at the classical level in field theory but broken by quantum effects. Fujikawa \cite{tre} realized that, at the quantum path integral level, the breaking of these symmetries comes from the non-invariance of the functional measure. 

At the end of the Eighties a path integral method was put forward also for {\it C}lassical {\it M}echanics (CM) \cite{quattro}. We will indicate it with the acronym CPI (for {\it C}lassical {\it P}ath {\it I}ntegral) to distinguish it from the {\it Q}uantum {\it P}ath {\it I}ntegral (which we will indicate with the acronym QPI). The reason to develop the CPI has been dictated mainly by the desire to better compare CM with QM. In fact, with both theories formulated via path integrals, their comparison has turned out to be very enlightening \cite{cinque}. In this paper we want first to extend the CPI to fermionic field theories along the lines pionereed for scalar and gauge fields in Ref. \cite{sei}. Second we want to study the chiral symmetry in the CPI formalism. As there cannot be any chiral anomaly at the classical level, we expect a ``cancellation" of it in the CPI. We use the word ``cancellation" because a part of the functional measure of the CPI is the same as the one of the QPI, so it generates a chiral anomaly. Nevertheless in the CPI measure there is another part made up of {\it auxiliary fields} which, we expect, should cancel the chiral anomaly. To check this mechanism we will use the method of Fujikawa \cite{tre} who derived, as mentioned before, the chiral anomaly from the non-invariance of the quantum measure. 

The paper is organized as follows: in Section 2 we will briefly review the CPI for the point particle \cite{quattro} and for the scalar fields \cite{sei}. In Section 3 we shall present the CPI for fermions that had never been worked out before in the literature. Section 4 is dedicated to the issue of the Lorentz covariance and of the Hermiticity condition used in the CPI for fermions. In Section 5, for completeness, we review the formalism of Fujikawa for the QPI while in Section 6 and 7 we apply it to the CPI proving in detail the cancellation of the chiral anomaly. 

\section{Review of the CPI for Point Particles and Scalar Fields}

Quantum mechanics (QM) has a well-known \cite{uno} path integral formulation (QPI) which can be summarized by the following formula
\begin{displaymath}
\displaystyle \langle q_f, t_f|q_{\scriptscriptstyle 0}, t_{\scriptscriptstyle 0} \rangle =\int {\mathscr D}^{\prime\prime}q \; \exp  \frac{i}{\hbar} S[q] . \label{II-1}
\end{displaymath}
It expresses the transition amplitude between an initial state $|q_{\scriptscriptstyle 0}, t_{\scriptscriptstyle 0}\rangle$ and a final one $\langle q_f, t_f|$ as ``the sum" over all paths between $(q_{\scriptscriptstyle 0}, t_{\scriptscriptstyle 0})$ and $(q_f, t_f)$ with weight given by the action $S[q]$. A more general object used in QM and QFT is the generating functional 
\begin{displaymath}
\displaystyle Z_{\scriptscriptstyle {\textrm{QM}}}[J]=\int {\mathscr D}q \; \exp\left[ \frac{i}{\hbar} \left(S[q]+\int dt \, J(t)q(t)\right)\right],
\label{II-2}
\end{displaymath}
where $J(t)$ is the external current. A natural question is whether an object like $Z_{\scriptscriptstyle{\textrm{QM}}}$ could be defined also for classical mechanics. The most natural answer is the following \cite{quattro}
\begin{equation}
\displaystyle Z_{\scriptscriptstyle {\textrm{CM}}}[J]=\int {\mathscr D}\varphi^a \, \widetilde{\delta}[\varphi^a-\varphi^a_{\textrm{cl}}] \, \exp\left[i\int dt \, J\varphi\right] \label{II-3}
\end{equation}
where $\varphi^a$ indicate the phase space coordinates:
\begin{displaymath}
\varphi^a=(q^1\cdots q^n,p^1 \cdots p^n), \qquad a=1, \ldots 2n
\end{displaymath}
and $\varphi^a_{\textrm{cl}}(t)$ indicates the solutions of the classical equations of motion
\begin{equation}
\dot{\varphi}^a=\omega^{ab}\partial_bH(\varphi) \label{II-4}
\end{equation}
with $H(\varphi)$ the Hamiltonian of the system and $\omega^{ab}$ the symplectic matrix. Basically $Z_{\scriptscriptstyle{\textrm {CM}}}[J]$ is the most natural generating functional for CM because it gives weight one to the classical path, i.e. the path which solves the classical equations of motion (\ref{II-4}), and zero to all the others.

We will show below that $Z_{\scriptscriptstyle{\textrm{CM}}}[J]$ is related to the operatorial approach to CM pionereed in the Thirties by Koopman and von Neumann \cite{sette}. In order to show that we have to perform some manipulations on (\ref{II-3}). Let us first rewrite the Dirac delta in (\ref{II-3}) as follows:
\begin{equation}
\displaystyle \widetilde{\delta}[\varphi^a-\varphi^a_{\textrm{cl}}]
=\widetilde{\delta}[\dot{\varphi}^a-\omega^{ab}\partial_bH] \, \textrm{det}\left(\delta^a_b\partial_t-\omega^{ac}\partial_c\partial_bH\right) \label{II-5}
\end{equation}
where $\widetilde{\delta}[\quad]$ is a functional Dirac delta and $\textrm{det}(\qquad )$ is a functional determinant. Next we Fourier transform the Dirac deltas on the RHS of (\ref{II-5}) as follows:
\begin{equation}
\widetilde{\delta}[\dot{\varphi}^a-\omega^{ab}\partial_bH]=
\int {\mathscr D}\lambda_a \, \exp\left[i\int dt \, \lambda_a[\dot{\varphi}^a-\omega^{ab}\partial_bH]\right] \label{II-6}
\end{equation}
where $\lambda_a$ is a set of $2n$ auxiliary variables. Introducing $4n$ Grassmann variables $(c^a,\bar{c}_a)$ we can rewrite the determinant in (\ref{II-5}) as
\begin{equation}
\displaystyle \textrm{det}(\delta_b^a\partial_t-\omega^{ac}\partial_c\partial_bH)
=\int {\mathscr D}\bar{c}_a{\mathscr D}c^a\, \exp\left[-\int dt \, \bar{c}_a[\delta_b^a\partial_t -\omega^{ac}\partial_c\partial_bH]c^b\right]. \label{II-7}
\end{equation}
Putting together (\ref{II-5}), (\ref{II-6}) and (\ref{II-7}) we can rewrite the generating functional $Z_{\scriptscriptstyle {\textrm{CM}}}[J]$ with zero external current as
\begin{equation}
Z_{\scriptscriptstyle {\textrm{CM}}}[0]= \int {\mathscr D}\varphi^a {\mathscr D}\lambda_a{\mathscr D}\bar{c}_a {\mathscr D}c^a \, \exp\left[i\int dt \, \widetilde{\cal L}\right]
\label{II-8}
\end{equation}
where
\begin{equation}
\displaystyle \widetilde{\cal L}=\lambda_a[\dot{\varphi}^a-\omega^{ab}\partial_bH]+i\bar{c}_a[\delta_b^a\partial_t-\omega^{ac}\partial_c\partial_bH]c^b. \label{II-9}
\end{equation}
The Hamiltonian associated to the $\widetilde{\cal L}$ of (\ref{II-9}) is
\begin{equation}
\displaystyle \widetilde{\cal H}=\lambda_a\omega^{ab}\partial_bH+i\bar{c}_a\omega^{ac}\partial_c\partial_bHc^b.
\label{II-10}
\end{equation}
Via the path integral (\ref{II-8}), we can define the concept of commutators as Feynman \cite{uno} did for quantum mechanics. The result is \cite{quattro}
\begin{equation}
\langle [\varphi^a,\varphi^b] \rangle=0, \qquad
\langle [\varphi^a, \lambda_b]\rangle =i\delta_b^a \label{II-11}, \qquad
\langle [c^a,\bar{c}_b] \rangle=\delta_b^a. 
\end{equation}
All the other commutators are zero.
The first commutator of (\ref{II-11}) indicates that $p$ and $q$ commute and this confirms that we are doing CM. From the second relation we get that a possible realization for $\lambda_a$ is the following one: 
\begin{equation}
\displaystyle \hat{\lambda}_a=-i\frac{\partial}{\partial \varphi^a}. \label{II-12}
\end{equation}
Substituting (\ref{II-12}) in the $\widetilde{\cal H}$ of (\ref{II-10}) we get for the non-Grassmannian part the following operator:
\begin{equation}
\hat{\widetilde{\cal H}}=-i \omega^{ab}\partial_bH\partial_a\equiv -i\hat{L} \label{II-13}
\end{equation}
where $\hat{L}$ is nothing else than the well-known Liouville operator of classical mechanics
\begin{equation}
\displaystyle \hat{L}=\frac{\partial H}{\partial p}\frac{\partial}{\partial q}-\frac{\partial H}{\partial q}\frac{\partial}{\partial p}. \label{Liouv}
\end{equation}
This is the operator of evolution that Koopman and von Neumann proposed in their operatorial formulation of CM \cite{sette}. Basically they introduced a Hilbert space of square integrable functions $\psi(\varphi)$ on phase space and proposed the following two postulates: 
\begin{itemize}
\item[1)] The evolution of the $\psi(\varphi)$ is given by the equation
\begin{displaymath}
\displaystyle \frac{\partial \psi}{\partial t}=-\hat{L}\psi; 
\end{displaymath}
\item[2)] The probability density $\rho(\varphi)$ of classical statistical mechanics can be obtained from the $\psi$ as 
\begin{displaymath}
|\psi(\varphi)|^2=\rho(\varphi).
\end{displaymath} 
\end{itemize}
From the postulates 1) and 2) it is easy to derive the classical Liouville equation for $\rho$:
\begin{displaymath}
\displaystyle \frac{\partial \rho}{\partial t}=-\hat{L}\rho. 
\end{displaymath}
All this and expecially formula (\ref{II-13}) confirms that our path integral formulation of CM (the CPI for short) is the functional counterpart of the Koopman and von Neumann operatorial approach to CM. 

The reader may anyhow object that, in making this connection, we have not taken into account the full $\widetilde{\cal H}$ but only its non-Grassmannian part. It is actually easy to prove that the full $\widetilde{\cal H}$ (with the Grassmann variables included) is a well-known generalization of the Liouville operator called the ``Lie derivative of the Hamiltonian flow" and often indicated with ${\cal L}_{(dH)^{\sharp}}$ \cite{otto}. In order to identify  $\widetilde{\cal H}$ with the Lie derivative the crucial step is to identify the Grassmann variables $c^a$ with the forms $d\varphi^a$ over the phase space manifold. Note that the anticommuting nature of the $c^a$ reproduces naturally the anticommuting feature of the wedge product ``$\wedge$" defined among differential forms. In our CPI the Lie derivative of the Hamiltonian flow
makes the evolution of the basis of the differential forms, $d\varphi^a$, and in general of any function $\rho(\varphi^a,d\varphi^a)$ of the phase space points and their associated forms \cite{quattro}. The general expression for $\rho(\varphi^a,d\varphi^a)$ is:
\begin{equation}
\rho(\varphi^a, d\varphi^a)=\rho_{\scriptscriptstyle 0}(\varphi) +\rho_a(\varphi)d\varphi^a+
\rho_{ab}(\varphi)d\varphi^a\wedge d\varphi^b+ \cdots. \label{II-17}
\end{equation}
The first term $\rho_{\scriptscriptstyle 0}(\varphi)$ in (\ref{II-17}) evolves with the Liouvillian $\hat{L}$ of (\ref{Liouv}) while the other terms need, for their evolution, an extra piece given by the Grassmann part contained in $\widetilde{\cal H}$. 

We want to stress that, while in the standard formulation of CM the time evolution is generated by the Hamiltonian $H$ via its Poisson brackets, in the CPI the same evolution is generated by the $\widetilde{\cal H}$ of Eq. (\ref{II-10}) via the commutators (\ref{II-11}). Analogously, while a function $O(\varphi)$ generates a canonical transformation via the Poisson brackets, in the operator approach to CM the same transformations are generated via the commutators (\ref{II-11}) by the Lie derivative along the Hamiltonian vector field associated with $O$ which is:
\begin{equation}
\widetilde{O}\equiv\lambda_a\omega^{ab}\partial_bO+i\bar{c}_a\omega^{ab}\partial_b\partial_dOc^d. \label{supero}
\end{equation}
This is also the rule to get conserved charges at the CPI level: if a charge $O(\varphi)$ is conserved in the standard formulation of CM then the associated $\widetilde{O}$ of Eq. (\ref{supero}) is conserved in the operator approach to CM, in the sense that it commutes with the $\widetilde{\cal H}$ of (\ref{II-10}).
We will not go on with further details but we refer the reader not familiar with this topic to the extensive literature on the subject \cite{quattro}. 

The formalism of the CPI can be easily extended to field theories as it has been done for scalar and gauge fields in Ref. \cite{sei}. We will here briefly review the case of scalar fields. Let us consider a $\varphi^4$ theory with Lagrangian density
\begin{displaymath}
\displaystyle {\cal L}=\frac{1}{2}(\partial_{\mu}\varphi)(\partial^{\mu}\varphi)
-\frac{1}{2}m^2\varphi^2-\frac{1}{4!}g\varphi^4. 
\end{displaymath}
Introducing the momenta $\displaystyle \pi\equiv \frac{\partial {\cal L}}{\partial \dot{\varphi}}$ we can easily perform the Legendre transform to get the Hamiltonian\footnote{From now on we shall indicate with $x=(\mathbf{x},t)$ the space-time variables and with ${\bf x}$ only the space ones. We shall also use Green letters for sums over space-time variables and Latin letters for sums over space variables only.} 
\begin{displaymath}
\displaystyle H=\int d{\bf x} \left[\frac{1}{2}\pi^2+\frac{1}{2}(\partial_i\varphi)^2+ \frac{1}{2}m^2\varphi^2+\frac{1}{4!}g\varphi^4\right]. 
\end{displaymath}
The equations of motion are 
\begin{displaymath}
\displaystyle \dot{\varphi}(x)=\frac{\partial H}{\partial \pi(x)}, \qquad \dot{\pi}(x)=-\frac{\partial H}{\partial \varphi(x)}
\end{displaymath}
where the derivatives $\partial$ must be intended as {\it functional derivatives}. Introducing the phase space variables $\xi^a(x)=(\varphi(x),\pi(x)), \; a=1,2$, the equations of motion can be written as
\begin{displaymath}
\displaystyle \dot{\xi}^a=\omega^{ab}\frac{\partial H}{\partial \xi^b}
\end{displaymath}
and the analog of the generating functional (\ref{II-8}) is
\begin{displaymath}
\displaystyle Z_{\scriptscriptstyle{\textrm{CM}}}^{\varphi^4}[0]\equiv \int {\mathscr D}\xi^a{\mathscr D}\lambda_a{\mathscr D}c^a{\mathscr D}\bar{c}_a \;
\exp\left[i\int dx \, \widetilde{\cal L}^{\varphi^4}\right]
\end{displaymath}
where
\begin{displaymath}
\widetilde{\cal L}^{\varphi^4}=\lambda_a\Biggl(\dot{\xi}^a-\omega^{ab}\frac{\partial H}{\partial \xi^b}\Biggr)+i\bar{c}_a\Biggl(\delta_b^a\partial_t-\omega^{ac}\frac{\partial^2H}{\partial \varphi^c\partial \varphi^b}\Biggr) c^b.
\end{displaymath}
Of course now the auxiliary variables are the fields $\lambda_a(x)$, $c^a(x)$, $\bar{c}_a(x)$ and the integration in the weight $\int dx \,\widetilde{\cal L}^{\varphi^4}$ is not just over the time $t$ as for the point particle but over the space-time $x$. All this is trivial but we thought it was useful to repeat it again in order to have all the exact integrations in place before going on to the case of fermions. 

\section{Classical Path Integral for Fermions}

In this section we will extend the CPI formalism to fermion fields. Let us start from the free fermion Lagrangian:
\begin{equation}
\displaystyle L=i\int d{\mathbf{x}}\,\bar{\psi}(x)\gamma^{\mu}
\partial_{\mu}\psi(x) \label{lagdirac}
\end{equation}
where $\psi(x)$ indicates four Grassmann fields $\psi_{\alpha}(x)$ transforming as a spinor under the Lorentz group and $\bar{\psi}$ is defined in the usual way as $\bar{\psi}=\psi^{\dagger}\gamma^{0}$. 
From the Lagrangian (\ref{lagdirac}) we can derive the 
following equations of motion for the fields $\psi$ and $\psi^{\dagger}$:
\begin{equation}
\dot{\psi}+\gamma^{0}\gamma^{l}\partial_{l}\psi=0, 
\qquad\quad\dot{\psi}^{\dagger}+\partial_{l}\psi^{\dagger}\gamma^{0}
\gamma^{l}=0. \label{Lageq}
\end{equation}
The momentum associated with the Grassmannian odd coordinate field
$\psi_{\alpha}(x)$ is defined as usual:
\begin{equation}
\displaystyle \pi_{\alpha}(x)\equiv\frac{\partial  L}{\partial\dot{\psi}_{\alpha}(x)}=
-i\psi^{\dagger}_{\alpha}(x). \label{momentum}
\end{equation} 
Consequently the fundamental Poisson brackets at equal times are:
\begin{equation}
\displaystyle \{\psi_{\alpha}(\mathbf{x},t),\psi^{\dagger}_{\beta}(\mathbf{y},t)\}=-i\delta_{\alpha\beta}
\delta(\mathbf{x}-\mathbf{y}). \label{Poisson}
\end{equation}
When we quantize the system by using the Dirac's correspondence rules 
$\{\cdot,\cdot\}\,\rightarrow\,-i[\cdot,\cdot]$, from (\ref{Poisson}) 
we get the following equal-time {\it quantum} anticommutator:
\begin{equation}
\displaystyle [\psi_{\alpha}(\mathbf{x},t),\psi^{\dagger}_{\beta}(\mathbf{y},t)]=\delta_{\alpha\beta}
\delta(\mathbf{x}-\mathbf{y}). \label{anticomm}
\end{equation}
The equations of motion (\ref{Lageq}) can also be obtained via the Poisson brackets (\ref{Poisson}) and the Hamiltonian associated to $L$ which is: 
\begin{equation}
\displaystyle H=-i\int d\mathbf{x} \;\psi^{\dagger}(x)\gamma^{0}
\gamma^{l}\partial_{l}\psi(x). \label{hamdirac}
\end{equation}
Their explicit form can be written as:
\begin{equation}
\displaystyle \dot{\psi}(x)=\{\psi(x),H\}=-i\frac{\partial H}{\partial 
\psi^{\dagger}(x)},\qquad \quad 
\dot{\psi}^{\dagger}(x)=\{\psi^{\dagger}(x),H\}=
-i\frac{\partial H}{\partial \psi(x)}. \label{2.5bis}
\end{equation}
If we put together in $\Psi^a\equiv(\psi,\psi^{\dagger})$
the coordinate field $\psi$ and $\psi^{\dagger}$, which is related to the momentum (\ref{momentum}), and if we define the $2\times 2$ symplectic matrix\footnote{Note that in this case the symplectic matrix is symmetric because the phase space fields are Grassmannian odd.}: 
$\displaystyle \omega^{ab}=\begin{pmatrix} 0 & 
1\cr 1 & 0
\end{pmatrix}$, the equations of motion (\ref{2.5bis}) can be rewritten as\footnote{The index ``$a$" runs from 1 to 2 and should not be confused with the Lorentz index $\alpha$ which runs from 1 to 4 and which we have omitted here for simplicity.}:
\begin{equation}
\displaystyle \dot{\Psi}^a(x)=\{\Psi^a(x),H\}=
-i\omega^{ab}\frac{\partial H}{\partial\Psi^b(x)}.
\label{compeq}
\end{equation}
This form of the equations of motion is very useful in order to implement the classical path integral 
procedure outlined in Section 2 for the point particle and the scalar field. In fact the analog for fermions of the {\it classical} generating functional presented in (\ref{II-3}) is:
\begin{displaymath}
Z_{\scriptscriptstyle \textrm{CM}}[0]=\int {\mathscr D}\Psi^a\,
\widetilde{\delta}[\Psi^a-\Psi^a_{\textrm{cl}}]
\end{displaymath}
where $\widetilde{\delta}$ indicates a functional Dirac delta and 
$\Psi^a_{\textrm{cl}}$ stands for 
the solutions to the equations (\ref{compeq}). If we pass from the Dirac delta of the solutions to the Dirac delta of the equations of motion, we get the following generating functional:
\begin{equation}
\displaystyle Z_{\scriptscriptstyle \textrm{CM}}= 
\int {\mathscr D}\Psi^a \,\widetilde{\delta}\biggl(\dot{\Psi}^a+i\omega^{ab}
\frac{\partial H}{\partial \Psi^b}\biggr)\textrm{det}^{-1}\biggl[\delta^a_d \partial_t
\delta(\mathbf{x}-\mathbf{y})+i\omega^{ab}\frac{\partial^2H}{\partial \Psi^d(\mathbf{y})
\partial\Psi^b(\mathbf{x})}\biggr]. \label{passage}
\end{equation}
Let us notice that in (\ref{passage}) there appears the inverse of a determinant instead of the determinant like in (\ref{II-5}), because the fields $\Psi^a$ are Grassmannian odd. 
If we use the explicit form (\ref{Lageq}) of the equations of motion
we can then rewrite Eq. (\ref{passage}) as:
\begin{displaymath}
\displaystyle Z_{\scriptscriptstyle \textrm{CM}}= \int {\mathscr D}\psi \,
\widetilde{\delta}\Bigl(\dot{\psi}+\gamma^{0}\gamma^{l}\partial_{l}\psi\Bigr)
\,\textrm{det}^{-1}\Bigl[\partial_t+\gamma^{0}\gamma^{l}\partial_{l}\Bigr]\int
{\mathscr D}\psi^{\dagger}\widetilde{\delta}\Bigl(\dot{\psi}^{\dagger}+
\partial_{l}\psi^{\dagger}\gamma^{0}\gamma^{l}\Bigr)
\textrm{det}^{-1}\Bigl[\partial_t+\partial_{l}\gamma^{0}\gamma^{l}\Bigr].
\end{displaymath} 
Similarly to what we did for the point particle, but taking into account the different Grassmannian character of the variables, let us now introduce the auxiliary Grassmannian odd fields
$\lambda_a=(\lambda_{\psi},\lambda_{\psi^{\dagger}})$ in order to exponentiate the equations of motion
and the Grassmannian even fields $c^a=(c^{\psi},c^{\psi^{\dagger}})$ and 
$\bar{c}_a=(\bar{c}_{\psi},\bar{c}_{\psi^{\dagger}})$ in order to write the inverse
of the determinants. The final result is
\begin{equation}
\displaystyle Z_{\scriptscriptstyle \textrm{CM}}[0]=
\int {\mathscr D}\Psi^a {\mathscr D}\lambda_a{\mathscr D}c^a{\mathscr
D}\bar{c}_a \,\textrm{exp} \biggl[
i\int dx \,\widetilde{\cal L}\biggr] \label{final}
\end{equation}
where $\widetilde{\cal L}$ is the following Lagrangian density:
\begin{eqnarray}
\displaystyle \label{suplag} \widetilde{\cal L}&=&
\lambda_{\psi}(\dot{\psi}+\gamma^{0}
\gamma^{l}\partial_{l}\psi)-(\dot{\psi}^{\dagger}+
\partial_{l}\psi^{\dagger}\gamma^{0}\gamma^{l})
\lambda_{\psi^{\dagger}}+i\bar{c}_{\psi}(\dot{c}^{\psi}+\gamma^{0}
\gamma^{l}\partial_{l}c^{\psi})+
i(\dot{c}^{\psi^{\dagger}}+\partial_{l}c^{\psi^{\dagger}}\gamma^{0}
\gamma^{l})\bar{c}_{\psi^{\dagger}}. \nonumber\\ \label{3dieci}
\end{eqnarray}
We have omitted for semplicity the Lorentz indices on all the fields of the theory. An important point to keep in mind is that, for the manner we have constructed Eq. (\ref{3dieci}), the fields $\psi$, $\lambda_{\psi^{\dagger}}$, $c^{\psi}$ and
$\bar{c}_{\psi^{\dagger}}$ turn out to be column vectors while $\psi^{\dagger},\lambda_{\psi},c^{\psi^{\dagger}}$
and $\bar{c}_{\psi}$ are row vectors. 

The Lagrangian density in (\ref{final}) $\widetilde{\cal L}$ can be split in a kinetic and a Hamiltonian part as follows:
\begin{equation}
\widetilde{\cal L}=\lambda_{\psi}\dot{\psi}-\dot{\psi}^{\dagger}
\lambda_{\psi^{\dagger}}+i\bar{c}_{\psi}\dot{c}^{\psi}+
i\dot{c}^{\psi^{\dagger}}\bar{c}_{\psi^{\dagger}}-\widetilde{\cal H}, \label{suplag3}
\end{equation}
where the Hamiltonian density $\widetilde{\cal H}$ is given by:
\begin{equation}
\displaystyle \widetilde{\cal H}=
-\lambda_{\psi}\gamma^{0}\gamma^{l}\partial_{l}\psi+
(\partial_{l}\psi^{\dagger})\gamma^{0}
\gamma^{l}\lambda_{\psi^{\dagger}}
-i\bar{c}_{\psi}\gamma^{0}\gamma^{l}
\partial_{l}c^{\psi}-i(\partial_{l}c^{\psi^{\dagger}})
\gamma^{0}\gamma^{l}\bar{c}_{\psi^{\dagger}}. \label{supham}
\end{equation}
The Lagrangian $\widetilde{\cal L}$ of (\ref{suplag3}) can also be written in a much more compact form as: 
\begin{equation}
\widetilde{\cal L}=\lambda_{\psi}\gamma^0\gamma^{\mu}\partial_{\mu}\psi-(\partial_{\mu}\psi^{\dagger})
\gamma^0\gamma^{\mu}\lambda_{\psi^{\dagger}}+i\bar{c}_{\psi}\gamma^0\gamma^{\mu}\partial_{\mu}c^{\psi}
+i(\partial_{\mu}c^{\psi^{\dagger}})\gamma^0\gamma^{\mu}\bar{c}_{\psi^{\dagger}}. \label{lagcomp}
\end{equation}

Like we did in Section 2 for the point particle, we can associate an operator formalism to the path integral approach described above. In particular we can derive from it, like we did in Eq. (\ref{II-11}) for the point particle, the graded commutators among all variables of the theory. The non-zero graded commutators for the fermion fields turn out to be: 
\begin{equation}
\begin{array}{l}
\Bigl[\psi_{\alpha}(\mathbf{x},t),\lambda_{\psi, \beta}(\mathbf{y},t)\Bigr]=
i\delta_{\alpha\beta}\delta(\mathbf{x}-\mathbf{y}), \medskip \\
\Bigl[c^{\psi}_{\alpha}(\mathbf{x},t),\bar{c}_{\psi,\beta}(\mathbf{y},t)\Bigr]=
\delta_{\alpha\beta}\delta(\mathbf{x}-\mathbf{y}),
\end{array} \begin{array}{l}
\Bigl[\psi^{\dagger}_{\alpha}(\mathbf{x},t),\lambda_{\psi^{\dagger},\beta}(\mathbf{y},t)\Bigr]=
i\delta_{\alpha\beta}\delta(\mathbf{x}-\mathbf{y}). \medskip\\
\Bigl[c^{\psi^{\dagger}}_{\alpha}(\mathbf{x},t),\bar{c}_{\psi^{\dagger},\beta}(\mathbf{y},t)\Bigr]=
\delta_{\alpha\beta}\delta(\mathbf{x}-\mathbf{y}). \label{supcomm}
\end{array}
\end{equation}
The equations of motion for the spinors $\Psi^a=(\psi,\psi^{\dagger})$ can be reproduced via the Hamiltonian $H$ of 
Eq. (\ref{hamdirac}) and the {\it quantum} anticommutator (\ref{anticomm}) as $\dot{\Psi}^a=i[\Psi^a,H]$ or via 
the Hamiltonian density (\ref{supham}) and the graded {\it classical} commutators (\ref{supcomm}) as $\dot{\Psi}^a=i[\Psi^a,\int d\mathbf{x}\,\widetilde{\cal H}]$. From $\widetilde{\cal L}$ or from $\widetilde{\cal H}$ and the commutators (\ref{supcomm})  
we can also derive the equations of motion for the auxiliary fields $\lambda_{\psi}$, $\lambda_{\psi^{\dagger}}$, $c^{\psi}$, $c^{\psi^{\dagger}}$, $\bar{c}_{\psi}$, $\bar{c}_{\psi^{\dagger}}$: 
\begin{displaymath}
\begin{array}{l}
\dot{\lambda}_{\psi}+\partial_l\lambda_{\psi}\gamma^0\gamma^l=0 \smallskip \\
\dot{c}^{\psi}+\gamma^0\gamma^l\partial_lc^{\psi}=0 \smallskip \\
\dot{\bar{c}}_{\psi}+\partial_l\bar{c}_{\psi}\gamma^0\gamma^l=0
\end{array} \qquad \begin{array}{l}
\dot{\lambda}_{\psi^{\dagger}}+\gamma^0\gamma^l\partial_l\lambda_{\psi^{\dagger}}=0
\smallskip \\
\dot{c}^{\psi^{\dagger}}+\partial_lc^{\psi^{\dagger}}\gamma^0\gamma^l=0 \smallskip \\
\dot{\bar{c}}_{\psi^{\dagger}}+\gamma^0\gamma^l\partial_l\bar{c}_{\psi^{\dagger}}=0.
\end{array} 
\end{displaymath}
An important issue that we should mention at this point is that, in implementing the CPI for fermions, 
we could have started also from the equations of motion for $\psi$ and $\bar{\psi}$ instead of the ones for $\psi$ and $\psi^{\dagger}$ as we did. The equations of motion for $\psi$ and $\bar{\psi}$ are
\begin{displaymath}
\dot{\psi}+\gamma^{0}\gamma^{l}\partial_{l}\psi=0\qquad\quad 
\dot{\bar{\psi}}+\partial_{l}\bar{\psi}\gamma^{l}\gamma^{0}=0. 
\end{displaymath}
This choice is the most suitable one if we want to compare 
the {\it quantum} and the {\it classical} path integrals. 
In fact the fields appearing in the functional measure of the 
{\it quantum} path integral are $\psi$ and 
$\bar{\psi}$ and the reason is that the measure ${\mathscr D}\psi{\mathscr D}\bar{\psi}$, differently from ${\mathscr D}\psi{\mathscr D}\psi^{\dagger}$, 
is manifestly Lorentz invariant. 
Since the equations of motion in $\psi$ and $\bar{\psi}$ are not coupled we can perform the same manipulations used before for $\psi$ and $\psi^{\dagger}$ in order to build the associated\footnote{We will put a bar on $\bar{Z}_{\scriptscriptstyle{\textrm{CM}}}$, $\bar{\cal L}$ and $\bar{\cal H}$ to indicate that in principle they are different than the analogous objects appearing respectively in (\ref{passage}), (\ref{suplag}) and (\ref{supham}).} CPI
\begin{eqnarray}
\displaystyle \bar{Z}_{\scriptscriptstyle \textrm{CM}}[0]&=&\int {\mathscr D}\psi{\mathscr D}\bar{\psi}\;
\widetilde{\delta}(\psi-\psi_{\textrm{cl}})\widetilde{\delta}(\bar{\psi}-\bar{\psi}_{\textrm{cl}})=\nonumber \\
&=& \int {\mathscr D}\psi\; \widetilde{\delta}(\dot{\psi}+\gamma^{0}
\gamma^{l}\partial_{l}\psi) \,\textrm{det}^{-1}(\partial_t+\gamma^{0}
\gamma^{l}\partial_{l}) \cdot \nonumber  \\
&& \cdot \int {\mathscr D}\bar{\psi} \; \widetilde{\delta}(\dot{\bar{\psi}}+\partial_{l}\bar{\psi}\gamma^{l}
\gamma^{0})\,
\textrm{det}^{-1}(\partial_t+\partial_{l}\gamma^{l}
\gamma^{0}). \nonumber
\end{eqnarray}
By exponentiating with the standard rules the Dirac deltas and the inverse of the functional determinants we get the following path integral:
\begin{equation}
\displaystyle \bar{Z}_{\scriptscriptstyle \textrm{CM}}[0]=\int {\mathscr D}\mu \,\textrm{exp}\biggl[i\int \, dx \,\bar{\cal L}\biggr]. \label{pathz}
\end{equation}
${\mathscr D}\mu$ stands for the integration over all the fields of the theory:
\begin{displaymath}
\mu=(\psi,\bar{\psi},\lambda_{\psi},\lambda_{\bar{\psi}},c^{\psi},c^{\bar{\psi}},\bar{c}_{\psi},\bar{c}_{\bar{\psi}}) \label{allv}
\end{displaymath}
and $\bar{\cal L}$ is the following Lagrangian density:
\begin{equation}
\bar{\cal L}=\lambda_{\psi}\dot{\psi}-\dot{\bar{\psi}}\lambda_{\bar{\psi}}+
i\bar{c}_{\psi}\dot{c}^{\psi}+i\dot{c}^{\bar{\psi}}\bar{c}_{\bar{\psi}}
-\bar{\cal H} \label{suplag2} 
\end{equation}
with 
\begin{displaymath}
\bar{\cal H}=-\lambda_{\psi}\gamma^{0}\gamma^{l}
\partial_{l}\psi+(\partial_{l}\bar{\psi})\gamma^{l}
\gamma^{0}\lambda_{\bar{\psi}}-i\bar{c}_{\psi}\gamma^{0}
\gamma^{l}\partial_{l}c^{\psi}
-i(\partial_{l}c^{\bar{\psi}})\gamma^{l}
\gamma^{0}\bar{c}_{\bar{\psi}}. 
\end{displaymath}
Like in (\ref{lagcomp}) the Lagrangian density $\bar{\cal L}$ can be written in a compact form as:
\begin{equation}
\displaystyle \bar{\cal L}=\lambda_{\psi}\gamma^0\gamma^{\mu}\partial_{\mu}\psi
-(\partial_{\mu}\bar{\psi})\gamma^{\mu}\gamma^0\lambda_{\bar{\psi}}
+i\bar{c}_{\psi}\gamma^0\gamma^{\mu}\partial_{\mu}c^{\psi}
+i(\partial_{\mu}c^{\bar{\psi}})\gamma^{\mu}\gamma^0\bar{c}_{\bar{\psi}}. \label{lagcomp2}
\end{equation}
From the Lagrangian density (\ref{suplag2}) or (\ref{lagcomp2})  we can derive the following equations of motion, which will be needed later on:
\begin{equation}
\begin{array}{l}
\dot{\psi}+\gamma^0\gamma^l\partial_l\psi=0 \smallskip \\
\dot{\lambda}_{\psi}+\partial_l\lambda_{\psi}\gamma^0\gamma^l=0 \smallskip \\
\dot{c}^{\psi}+\gamma^0\gamma^l\partial_lc^{\psi}=0 \smallskip \\
\dot{\bar{c}}_{\psi}+\partial_l\bar{c}_{\psi}\gamma^0\gamma^l=0
\end{array} \qquad \begin{array}{l}
\dot{\bar{\psi}}+\partial_l\bar{\psi}\gamma^l\gamma^0=0\smallskip \\
\dot{\lambda}_{\bar{\psi}}+\gamma^l\gamma^0\partial_l\lambda_{\bar{\psi}}=0\smallskip\\
\dot{c}^{\bar{\psi}}+\partial_lc^{\bar{\psi}}\gamma^l\gamma^0=0 \smallskip\\
\dot{\bar{c}}_{\bar{\psi}}+\gamma^l\gamma^0\partial_l\bar{c}_{\bar{\psi}}=0. \label{eqbar}
\end{array}
\end{equation}
The non-zero equal-time graded commutators
which can be derived from the path integral (\ref{pathz}) are the following ones:
\begin{equation}
\begin{array}{l}
\Bigl[\psi_{\alpha}(\mathbf{x},t),\lambda_{\psi, \beta}(\mathbf{y},t)\Bigr]=
i\delta_{\alpha\beta}\delta(\mathbf{x}-\mathbf{y}), \medskip  \\
\Bigl[c^{\psi}_{\alpha}(\mathbf{x},t),\bar{c}_{\psi, \beta}(\mathbf{y},t)\Bigr]=
\delta_{\alpha\beta}\delta(\mathbf{x}-\mathbf{y}) \end{array} \quad 
\begin{array}{l}
\Bigl[\bar{\psi}_{\alpha}(\mathbf{x},t),\lambda_{\bar{\psi}, \beta}(\mathbf{y},t)\Bigr]=
i\delta_{\alpha\beta}\delta(\mathbf{x}-\mathbf{y}), \medskip \\
\Bigl[c^{\bar{\psi}}_{\alpha}(\mathbf{x},t),\bar{c}_{\bar{\psi}, \beta}(\mathbf{y},t)\Bigr]
=\delta_{\alpha\beta}\delta(\mathbf{x}-\mathbf{y}). \label{supcomm2}
\end{array}
\end{equation}
The reader has for sure realized that the auxiliary variables entering the Lagrangian $\bar{\cal L}$ of (\ref{suplag2}) are different than those entering the $\widetilde{\cal L}$ of (\ref{suplag}). While the basic variables $(\psi,\bar{\psi})$ and $(\psi,\psi^{\dagger})$ are linked by the standard rule $\bar{\psi}=\psi^{\dagger}\gamma^0$, the rules linking the auxiliary variables of the CPI are: 
\begin{equation}
\bar{\psi}=\psi^{\dagger}\gamma^{0},\qquad \lambda_{\bar{\psi}}=
\gamma^{0}\lambda_{\psi^{\dagger}}, \qquad c^{\bar{\psi}}=c^{\psi^{\dagger}}
\gamma^{0}, \qquad \bar{c}_{\bar{\psi}}=\gamma^{0}\bar{c}_{\psi^{\dagger}}. \label{rul}
\end{equation}
It is in fact easy to check that, using (\ref{rul}) in the $\bar{\cal L}$ of (\ref{suplag2}) and in the commutators (\ref{supcomm2}), we obtain exactly the $\widetilde{\cal L}$ of (\ref{suplag}) and the commutators of (\ref{supcomm}). 

\section{Scalar Products and Lorentz Covariance}

In this section we want to analyze the issue of the Lorentz covariance and the choice of the scalar product of the field theory we are studying in this paper. First of all let us stress again that the functional measure of the quantum path integral \cite{tre} contains the two spinors $\psi$ and $\bar{\psi}$ instead of $\psi$ and $\psi^{\dagger}$ just because the measure ${\mathscr D}\psi{\mathscr D}\bar{\psi}$ is Lorentz invariant. In fact, let us indicate with $S$ the $4\times 4$ matrix that implements the Lorentz transformation of the spinor $\psi$, \cite{dieci}:
\begin{equation}
\psi^{\prime}(x^{\prime})=S(a)\psi(x). \label{numera1}
\end{equation}
The coordinates $x^{\prime}$ and $x$ are related as: $x^{\prime\,\nu}=a^{\nu}_{\mu}x^{\mu}$ with $a^{\nu}_{\mu}$ the $4\times 4$ matrix representing the particular Lorentz transformation that we perform. 
One can then deduce from (\ref{numera1}) that the Lorentz transformation of the adjoint spinor $\bar{\psi}$ is given by:
\begin{displaymath}
\bar{\psi}^{\prime}(x^{\, \prime})=\bar{\psi}(x)S^{-1}. 
\end{displaymath}
Since $\psi$ and $\bar{\psi}$ transform with inverse matrices under Lorentz transformations, it is easy to realize that both the Lagrangian 
$L=i \int d{\mathbf x}\,\bar{\psi}\gamma^{\mu}\partial_{\mu}\psi$ and the functional measure ${\mathscr D}\psi{\mathscr D}\bar{\psi}$ are explicitly Lorentz invariant, differently from what happens in the case of the measure ${\mathscr D}\psi{\mathscr D}\psi^{\dagger}$. 

Let us now analyze the operator formalism associated with 
the classical path integral (\ref{pathz}) and in particular let us choose a scalar product in the Hilbert space underlying the CPI for fermions. 
This last issue is a non trivial one when Grassmann variables are involved. In fact, as we proved in Ref. \cite{nove}, in a Hilbert space with Grassmann variables one can introduce several scalar products but the only one {\it positive definite} is the one for which the momentum conjugate to each Grassmann variable is related to the Hermitian conjugate of that same variable\footnote{This is for example what happens in the QFT of fermions where the only anticommutator different from zero is
$[\psi_{\alpha}({\bf x}),\psi_{\beta}^{\dagger}({\bf y})]=\delta_{\alpha \beta}\delta({\bf x}-{\bf y})$.
This choice of the scalar product guarantees the Hermiticity of the $H$ of Eq. (\ref{hamdirac}).}, i.e. $[c^a, c^{b\dagger}]=\delta^{ab}$.
If we limit ourselves to the 
Grassmann part of the theory, the equal-time anticommutators (\ref{supcomm2})
associated with the Lagrangian density\footnote{The index $g$ indicates that we have considered only the Grassmann part of the theory.}
$\bar{\cal L}_g=\lambda_{\psi}\dot{\psi}-\dot{\bar{\psi}}\lambda_{\bar{\psi}}-\bar{\cal H}_g$ can be rewritten as:
\begin{displaymath}
\displaystyle [\psi_{\alpha}(\mathbf{x}),-i\lambda_{\psi,\beta}(\mathbf{y})]=\delta_{\alpha\beta}
\delta(\mathbf{x}-\mathbf{y}),\qquad\quad [\lambda_{\bar{\psi},\beta}(\mathbf{x}),
-i\bar{\psi}_{\alpha}(\mathbf{y})]=\delta_{\alpha\beta}
\delta(\mathbf{x}-\mathbf{y}).
\end{displaymath}
If we look at the column vectors $\psi_{\alpha}$ and $\lambda_{\bar{\psi}, \beta}$ as
coordinates, then, as explained above and more in detail in Ref. \cite{nove}, in order to have a positive definite scalar product
the canonically conjugated momenta $-i\lambda_{\psi, \beta}$ and $-i\bar{\psi}_{\alpha}$
must be the Hermitian conjugate of $\psi_{\beta}$ and $\lambda_{\bar{\psi}, \alpha}$ respectively, i.e.:
\begin{equation}
-i\lambda_{\psi, \beta}=
\psi_{\beta}^{\dagger} \qquad \quad
-i\bar{\psi}_{\alpha}=(\lambda_{\bar{\psi}, \alpha})^{\dagger}. \label{herm}
\end{equation}

As a side observation let us notice that 
the analysis on the scalar products above and in particular Eq. (\ref{herm}) turns out to be useful in the study of the Lorentz covariance of the CPI (\ref{pathz}) written in terms 
of the fields $\psi$ and $\bar{\psi}$. In fact if $\psi$ and $\bar{\psi}$ transform under a 
Lorentz transformation in the usual way, i.e. $\psi^{\prime}=S\psi$ and 
$\bar{\psi}^{\prime}=\bar{\psi}S^{-1}$, then
the way the variables $\lambda$ transform can be derived by the particular scalar product imposed in (\ref{herm}):
\begin{equation}
\left\{
\begin{array}{l}
\lambda_{\psi}^{\prime}(x^{\prime})= i\bigl[\psi^{\prime}(x^{\prime})\bigr]^{\dagger}=
i\bigl[S\psi(x)\bigr]^{\dagger}=i\psi^{\dagger}(x)S^{\dagger}=
\lambda_{\psi}(x)S^{\dagger} \label{4-4} \medskip\\
\lambda_{\bar{\psi}}^{\prime}(x^{\prime})=i\bigl[\bar{\psi}^{\prime}(x^{\prime})\bigr]^{\dagger}=
i\bigl[\bar{\psi}(x)S^{-1}\bigr]^{\dagger}=i(S^{-1})^{\dagger}\bar{\psi}^{\dagger}(x)=
(S^{\dagger})^{-1}\lambda_{\bar{\psi}}(x).
\end{array}
\right.
\end{equation}
This is due to the fact that we want the Hermiticity relations (\ref{herm}) to be covariant under Lorentz transformations\footnote{In Appendix A we will prove that, starting from $\lambda_{\psi}$ and $\lambda_{\bar{\psi}}$, it is possible to build two new fields $\lambda$ and $\bar{\lambda}$ transforming with $S$ and $S^{-1}$ respectively and we will implement the associated CPI.}. 
From Eq. (\ref{4-4}) we deduce that, besides ${\mathscr D}\psi{\mathscr D}\bar{\psi}$, also the functional measure $\mathscr{D}\lambda_{\psi}{\mathscr D}\lambda_{\bar{\psi}}$ is invariant under Lorentz transformations. As a consequence, the whole CPI measure ${\mathscr D}\psi{\mathscr D}\bar{\psi}{\mathscr D}\lambda_{\psi}
{\mathscr D}\lambda_{\bar{\psi}}$ is invariant.
This analysis tells us that, if we had not chosen the Hermiticity conditions of Eq. (\ref{herm}), the Lorentz invariance of the CPI measure would not have been guaranteed. 

It is easy to prove that also the Grassmann part of the 
Lagrangian (\ref{lagcomp2}), which appears in the weight of the CPI (\ref{pathz}), is invariant under Lorentz transformations. In fact:
\begin{eqnarray}
\bar{\cal L}^{\prime}_g&=&\lambda_{\psi}^{\prime}\gamma^{0}
\gamma^{\mu}\partial_{\mu}^{\,\prime}\psi^{\prime}-(\partial_{\,\mu}^{\prime}\bar{\psi}^{\prime})\gamma^{\mu}
\gamma^{0}\lambda_{\bar{\psi}}^{\prime}=\nonumber \\ 
&=& \lambda_{\psi}S^{\dagger}\gamma^{0}
\gamma^{\mu}(\alpha^{-1})^{\nu}_{\mu}\partial_{\nu}(S\psi)
-(\alpha^{-1})^{\nu}_{\mu}\partial_{\nu}(\bar{\psi}S^{-1})\gamma^{\mu}\gamma^{0}
(S^{-1})^{\dagger}\lambda_{\bar{\psi}}=\nonumber\\ &=&\lambda_{\psi}\gamma^{0}S^{-1}
\gamma^{\mu}(\alpha^{-1})^{\nu}_{\mu}S\partial_{\nu} \psi
-(\alpha^{-1})^{\nu}_{\mu}(\partial_{\nu}\bar{\psi})S^{-1}\gamma^{\mu}S\gamma^0\lambda_{\bar{\psi}}=\nonumber\\
&=& \lambda_{\psi}\gamma^{0}\gamma^{\nu}\partial_{\nu}\psi-(\partial_{\nu}\bar{\psi})
\gamma^{\nu}\gamma^{0}\lambda_{\bar{\psi}}=\bar{\cal L}_g, \nonumber
\end{eqnarray}
where we have used the following properties of the matrices $S$ and $\gamma$:
\begin{equation}
\begin{array}{c}
\gamma^{0} S^{\dagger} \gamma^{0}=S^{-1}\; \Longrightarrow \; S^{\dagger}\gamma^{0}=\gamma^{0}S^{-1}
\; \Longrightarrow \; \gamma^0 (S^{-1})^{\dagger}=S\gamma^0 \medskip \\
S^{-1}\gamma^{\mu}S=\alpha^{\mu}_{\nu}\gamma^{\nu}. \label{propesse}
\end{array}
\end{equation}
This proves that not only the Grassmann part of the functional measure,
but also the Grassmann part of the Lagrangian density $\bar{\cal L}$ of Eq. (\ref{lagcomp2}) is Lorentz invariant. 

The behavior under Lorentz transformations of the auxiliary variables $(c,\bar{c})$, used to exponentiate the inverse of the determinant, cannot instead be determined with the same strategy used before. The reasons are basically two: first of all, we do not know a priori any Lorentz transformation equation for these auxiliary variables;
second, since the variables $c$ and $\bar{c}$ are Grassmannian even, it is not necessary to connect them via an operation of Hermitian conjugate to have a positive definite scalar product, like in the Grassmannian odd case. To understand better this last point we invite the reader to throughly study Ref. \cite{nove}. 
The strategy that we will adopt is the following one: as the variables
$c^{\psi}$ and $c^{\bar{\psi}}$ evolve in time as the first variations \cite{quattro} $\delta\psi$ and $\delta\bar{\psi}$ of the fields $\psi$, $\bar{\psi}$, see Eq. (\ref{eqbar}), we will require that they also transform under Lorentz like $\psi$ and $\bar{\psi}$ do:
\begin{equation}
c^{\psi^{\prime}}=Sc^{\psi}, \qquad\;\; c^{\bar{\psi}^{\,\prime}}=c^{\bar{\psi}}S^{-1}. \label{ciuno}
\end{equation}
The requirement of relativistic covariance of the theory will then uniquely determine the
Lorentz transformations of the other set of auxiliary variables, i.e. the $\bar{c}$. In fact, 
the part of the Lagrangian that depends on the $c$, $\bar{c}$ variables is:
\begin{displaymath}
\bar{\cal L}_{c,\bar{c}}=i\bar{c}_{\psi}\gamma^0\gamma^{\mu}\partial_{\mu}
c^{\psi}+i(\partial_{\mu}c^{\bar{\psi}})\gamma^{\mu}\gamma^0\bar{c}_{\bar{\psi}}
\end{displaymath}
and it has the same structure of the Grassmann part:
\begin{displaymath}
\bar{\cal L}_g=\lambda_{\psi}\gamma^0
\gamma^{\mu}\partial_{\mu}\psi-(\partial_{\mu}\bar{\psi})
\gamma^{\mu}\gamma^0\lambda_{\bar{\psi}}.
\end{displaymath}
Consequently the covariance of the theory under Lorentz transformations will be guaranteed, if we use (\ref{ciuno}), by requiring that $\bar{c}_{\psi}$ and $\bar{c}_{\bar{\psi}}$ transform just as $\lambda_{\psi}$ and $\lambda_{\bar{\psi}}$ respectively:
\begin{equation}
\bar{c}_{\psi}^{\,\prime}=\bar{c}_{\psi}S^{\dagger}, \qquad \bar{c}_{\bar{\psi}}^{\,\prime}=(S^{\dagger})^{-1}\bar{c}_{\bar{\psi}}. \label{cidue}
\end{equation}
Note that the transformations (\ref{ciuno}) and (\ref{cidue}) not only guarantee the Lorentz invariance of the $\bar{\cal L}_{c,\bar{c}}$ but also of the associated measure
${\mathscr D}c^{\psi}{\mathscr D}c^{\bar{\psi}}{\mathscr D}\bar{c}_{\psi}{\mathscr D}\bar{c}_{\bar{\psi}}$. So we can conclude that the overall measure
${\mathscr D}\psi{\mathscr D}\bar{\psi}{\mathscr D}\lambda_{\psi}
{\mathscr D}\lambda_{\bar{\psi}}
{\mathscr D}c^{\psi}{\mathscr D}c^{\bar{\psi}}{\mathscr D}\bar{c}_{\psi}{\mathscr D}\bar{c}_{\bar{\psi}}$
and the full Lagrangian $\bar{\cal L}$ are Lorentz invariant.

In Section 3 we had built another CPI: the one in the fields $\psi$ and $\psi^{\dagger}$. A natural question is whether this path integral is invariant or not under Lorentz transformations. At the quantum level the answer was no but at the CPI level the answer unexpectedly is yes and the details are given in Appendix B. 


\section{Chiral Symmetry and Quantum Path Integral}

In the previous sections we have studied a free massless fermion theory. In this section we will put the massless fermions in interaction with a $U(1)$ gauge field and we will briefly review the Fujikawa method for the evaluation of the chiral anomaly at the quantum level. The Lagrangian of the interacting theory is:
\begin{equation}
\displaystyle {\mathscr L}=i\bar{\psi}\gamma^{\mu}D_{\mu}\psi-\frac{1}{4}
F_{\mu \nu}F^{\mu \nu} \label{lagfuj}
\end{equation}
where $D_{\mu}$ stands for the covariant derivative $D_{\mu}=\partial_{\mu}+ieA_{\mu}$.
It is easy to check that the Lagrangian density ${\mathscr L}$ is invariant under the infinitesimal transformations:
\begin{equation}
\begin{array}{l}
\displaystyle \psi({x})\,\longrightarrow \,\textrm{exp}\,[i\alpha \gamma^5]\,
\psi({x})\; \simeq \;[1+i\alpha\gamma^5 ]\psi({x}), \medskip\\
\displaystyle \bar{\psi}({x})\,\longrightarrow \, \bar{\psi}({x})\,\textrm{exp}\,
[i\alpha\gamma^5] \; \simeq \; \bar{\psi}({x})[1+i\alpha\gamma^5].
\label{chirtra}
\end{array}
\end{equation}
This is what is known as the chiral symmetry of the theory.
The associated current and conserved charge are 
\begin{equation}
J^{\mu}_5=\bar{\psi}\gamma^{\mu}\gamma^5\psi \label{noether}
\end{equation}
and
\begin{equation}
\displaystyle Q_5=\int d\mathbf{x}\,\psi^{\dagger}(x)\gamma^5\psi(x). \label{qu5}
\end{equation}
Using the quantum anticommutator 
\begin{displaymath}
\displaystyle [\psi_{\alpha}(\mathbf{x},t),\psi^{\dagger}_{\beta}(\mathbf{y},t)]=\delta_{\alpha\beta}
\delta(\mathbf{x}-\mathbf{y}). 
\end{displaymath} 
and the rules 
\begin{displaymath}
\delta \psi=i\alpha [\psi,Q_5], \qquad \qquad \delta \bar{\psi}=
i\alpha [\bar{\psi},Q_5],
\end{displaymath}
it is easy to show that $Q_5$ generates just the infinitesimal chiral transformations (\ref{chirtra}). 
 
In quantizing the system via the path integral approach one gets the following object:
\begin{displaymath}
\displaystyle Z_{\scriptscriptstyle \textrm{QM}}[0]=
\int [{\mathscr D}A_{\mu}(x)]{\mathscr D}\bar{\psi}(x)
{\mathscr D}\psi(x)\,\textrm{exp} \,\biggl[\frac{i}{\hbar}\int dx \,{\mathscr L}\biggr]
\end{displaymath} 
which is the generating functional with the external currents put to zero. In $Z_{\scriptscriptstyle {\textrm{QM}}}$ we indicated formally with $[{\mathscr D}A_{\mu}]$ the integration over the gauge fields $A_{\mu}$ plus the Faddeev-Popov ghosts and determinant. As in $Z_{\scriptscriptstyle {\textrm{QM}}}[0]$ the fields are integrated over, we can perform any change of variables without modifying the generating functional. In particular, if we implement the following {\it local} version of the chiral transformations:
\begin{equation}
\begin{array}{l}
\displaystyle \psi({x})\,\longrightarrow \,\textrm{exp}\,[i\alpha(x) \gamma^5]\,
\psi({x})\; \simeq \;[1+i\alpha(x)\gamma^5 ]\psi({x}), \medskip\\
\displaystyle \bar{\psi}({x})\,\longrightarrow \, \bar{\psi}({x})\,\textrm{exp}\,
[i\alpha(x)\gamma^5] \; \simeq \; \bar{\psi}({x})[1+i\alpha(x)\gamma^5]
\label{chirtraloc}
\end{array}
\end{equation}
we get that the Lagrangian changes as
\begin{displaymath}
{\mathscr L} \; \longrightarrow \; {\mathscr L} -\partial_{\mu}\alpha(x)J_5^{\mu}(x)
\end{displaymath}
and $Z_{\scriptscriptstyle{\textrm{QM}}}$ becomes:
\begin{displaymath}
\displaystyle Z_{\scriptscriptstyle \textrm{QM}}=\int [{\mathscr D}A_{\mu}(x)]
{\mathscr D}\bar{\psi}^{\prime}(x){\mathscr D}\psi^{\prime}(x)
\,\textrm{exp} \biggl[\frac{i}{\hbar}\int dx \, [{\mathscr L}+\alpha(x)\partial_{\mu}J_5^{\mu}]\biggr],
\end{displaymath}
where $J_5^{\mu}$ is the Noether current of Eq. (\ref{noether}). As the generating functional $Z_{\scriptscriptstyle \textrm{QM}}$ must be unchanged, it cannot depend on the parameter $\alpha(x)$, i.e.:
\begin{equation}
\displaystyle \frac{\partial Z_{\scriptscriptstyle \textrm{QM}}}{\partial \alpha(x)}\biggl|_{\alpha=0}=0.
\label{independ}
\end{equation}
If the functional measure were invariant under the transformations (\ref{chirtraloc}) 
then Eq. (\ref{independ}) would give:
\begin{displaymath}
\displaystyle \int [{\mathscr D}A_{\mu}]{\mathscr D}\bar{\psi}{\mathscr D}\psi \;
(\partial_{\mu}J_5^{\mu}) \exp \left[{\frac{i}{\hbar}\int dx{\mathscr L}}\right]=0 
\end{displaymath}
which means:
\begin{equation}
\langle \partial_{\mu}J_5^{\mu}\rangle=0, \label{1.y}
\end{equation}
i.e., we would get the conservation law at the quantum mechanical level.
In order to obtain the result (\ref{1.y}) it was crucial to assume that the measure were invariant under the chiral transformations.
This is not the case for this model \cite{tre}. In fact, the quantum functional measure is not invariant but it transforms with a Jacobian $J$ different from one \cite{Bertlemann}:
\begin{equation}
\displaystyle {\mathscr D}\bar{\psi}^{\prime}{\mathscr D}\psi^{\prime}=
J\,{\mathscr D}\bar{\psi}{\mathscr D}\psi\equiv \exp \biggl[ i\int dx\,\alpha(x){\mathscr A}[A_{\mu}](x)\biggr]\,
{\mathscr D}\bar{\psi}{\mathscr D}\psi. \label{jacexp}
\end{equation}
In (\ref{jacexp}) we have rewritten the Jacobian $J$ as the exponential of a functional ${\mathscr A}$ which depends on the gauge fields $A_{\mu}$. In this case the requirement (\ref{independ}) that the functional integral does not depend on the parameter $\alpha(x)$ would produce 
\begin{displaymath}
\langle \partial_{\mu}J_5^{\mu}\rangle =\hbar \langle {\mathscr A}\rangle
\end{displaymath} 
instead of Eq. (\ref{1.y}). 
This would mean that at the QM level the chiral current $J_5^{\mu}$ for the model (\ref{lagfuj}) is no longer conserved.
As we said in the Introduction, Fujikawa \cite{tre} was the first to calculate anomalies evaluating explicitly the Jacobian $J$. For completeness, in the rest of this section we will review his method. 

Consider a Hermitian operator $\hat{A}$ with a complete set of eigenstates 
$\{\phi_n({x})\}$ defined by the following eigenvalue equation:
\begin{equation}
\hat{A}\phi_n({x})=\lambda_n\phi_n({x}). \label{eig}
\end{equation}
The Hermitian conjugate of (\ref{eig}) gives:
\begin{displaymath}
\phi_n^{\dagger}({x})\hat{A}=\lambda_n\phi_n^{\dagger}({x}).
\end{displaymath}
The states $\phi_n({x})$ satisfy the following orthonormal and completeness relations:
\begin{displaymath}
\displaystyle \int d{x}\,\phi_m^{\dagger}({x})\phi_n({x})=\delta_{mn},\qquad
\displaystyle \sum_n\phi_n({y})\phi_n^{\dagger}({x})=\delta({x}-{y}). 
\end{displaymath}
Let us suppose the states $\{\phi_n({x})\}$ make up a basis for the space of fermion fields. It is then possible to expand $\psi$ and $\bar{\psi}$ on the basis given by $\phi_n(x)$ and $\phi^{\dagger}_n(x)$ respectively:
\begin{displaymath}
\displaystyle \psi({x})=\sum_nb_n\phi_n({x}), \qquad
\bar{\psi}({x})=\sum_n\phi_n^{\dagger}({x})\bar{b}_n, 
\end{displaymath}
where $b_n$ and $\bar{b}_n$ are Grassmannian odd coefficients. As a result of this expansion, the functional measure can be rewritten as $\displaystyle {\mathscr D}\bar{\psi}{\mathscr D}\psi=\prod_n d\bar{b}_ndb_n$. We will prove that this measure is not invariant under a chiral transformation,
but it generates a Jacobian $J$ that can be evaluated explicitly \cite{tre}. Since the functional measure has been reduced to a product of numerical integrations over the coefficients $b_n$ and $\bar{b}_n$, 
it will be sufficient to calculate the transformation matrix of such coefficients. 
Let us start with the field $\psi$. From the infinitesimal chiral transformation:
\begin{displaymath}
\displaystyle \psi^{\prime}({x})=\biggl[1+i\alpha({x})\gamma^5\biggr]\psi({x})
\;\; \Longrightarrow \;\; \sum_nb_n^{\prime}\phi_n({x})
=\biggl[1+i\alpha({x})\gamma^5\biggr]\sum_nb_n\phi_n({x})
\end{displaymath}
we get:
\begin{equation}
\displaystyle b_m^{\prime}=\sum_n\int d{x} \,\phi_m^{\dagger}({x})
\bigl[1+i\alpha({x})\gamma^5\bigr]\phi_n({x}) \, b_n=\sum_n C_{mn} \; b_n, \label{numera13}
\end{equation}
where $C_{mn}$ is the following matrix:
\begin{equation}
\displaystyle C_{mn}\equiv \delta_{mn}+i\int d{x}\, \alpha({x})\phi_m^{\dagger}({x})
\gamma^5\phi_n({x}). \label{matrixc}
\end{equation}
From (\ref{numera13}) we have that the integration measure over the coefficients $b_n$ changes as follows \cite{Bertlemann}:
\begin{equation}
\prod_n db^{\prime}_n=[\textrm{det}\, C]^{-1}\prod_n db_n. \label{transf}
\end{equation}
In Eq. (\ref{transf}) there appears the inverse of a determinant because 
the coefficients $b_n$ are Grassmannian odd. With a similar procedure 
we get the following transformation for the coefficients $\bar{b}_n$:
\begin{displaymath}
\displaystyle \prod_n d\bar{b}_n^{\prime}=[\textrm{det}\, C]^{-1}\prod_nd\bar{b}_n.
\end{displaymath}
By putting together all the previous results we have that the functional measure transforms as follows:
\begin{equation}
{\mathscr D}\bar{\psi}^{\prime}{\mathscr D}\psi^{\prime}=
[\textrm{det}\,C]^{-2}{\mathscr D}\bar{\psi}{\mathscr D}\psi. \label{quant}
\end{equation}
This is the way the quantum functional measure changes under an infinitesimal chiral transformation.
The functional determinant appearing in (\ref{quant}) can be evaluated using a suitable regularization procedure, which preserves the gauge invariance of the theory \cite{tre}. Such procedure reproduces exactly the chiral anomaly that was first obtained via perturbative methods by Adler, Bell and Jackiw in \cite{due}.

\section{Chiral Symmetry at the Classical Level}

In this section we will turn to the CPI for the model given by the Lagrangian (\ref{lagfuj}). We know that at the classical level there is no anomaly, so we must expect a cancellation of the various Fujikawa Jacobians arising from the transformation of the CPI functional measure. This is what we will check explicitly in this and the next section. 

If we include the interaction with a gauge field $A_{\mu}$ like in (\ref{lagfuj}), the only change in the
equations of motion (\ref{eqbar}) for the free spinors $\psi$ and $\bar{\psi}$ is given by the replacement of the derivative $\partial_{\mu}$ with the covariant derivative $D_{\mu}=\partial_{\mu}+ieA_{\mu}$:
\begin{displaymath}
\dot{\psi}+\gamma^0\gamma^l\partial_l\psi+ie\gamma^0\gamma^{\mu}A_{\mu}\psi=0,\qquad
\dot{\bar{\psi}}+\partial_l\bar{\psi}\gamma^l\gamma^0+ieA_{\mu}\bar{\psi}\gamma^{\mu}\gamma^0=0.
\end{displaymath}
The same replacement must be made in the Lagrangian densities (\ref{lagcomp})-(\ref{lagcomp2}). So the CPI Lagrangian describing the interaction of the fermions with the gauge field becomes:
\begin{equation}
\widetilde{\cal L}=\lambda_{\psi}\gamma^0\gamma^{\mu}D_{\mu}\psi-(D_{\mu}\psi^{\dagger})\gamma^0\gamma^{\mu}\lambda_{\psi^{\dagger}}+i\bar{c}_{\psi}\gamma^0\gamma^{\mu}D_{\mu}c^{\psi}+i(D_{\mu}c^{\psi^{\dagger}})\gamma^0\gamma^{\mu}\bar{c}_{\psi^{\dagger}}. \label{6-1}
\end{equation}
The CPI functional measure must be extended to include also the CPI terms needed to describe the gauge field $A_{\mu}$. Those terms would appear in the analog of the functional measure $[{\mathscr D}A_{\mu}]$ of Fujikawa's quantum path integral 
but, since they do not change under chiral transformations, they will not play any role in the study of the chiral anomaly at the classical level. 

The Lagrangian (\ref{6-1}) is invariant under the following set of transformations: 
\begin{eqnarray}
\delta \psi=i\alpha\gamma^5\psi, &&\quad \delta\psi^{\dagger}=-i\alpha \psi^{\dagger}\gamma^5,
\nonumber\\
\delta \lambda_{\psi}=-i\alpha \lambda_{\psi}\gamma^5, &&\quad \delta \lambda_{\psi^{\dagger}}
=i\alpha \gamma^5 \lambda_{\psi^{\dagger}}, \label{suptr}\\
\delta c^{\psi}=i\alpha \gamma^5c^{\psi}, &&\quad \delta c^{\psi^{\dagger}}=
-i\alpha c^{\psi^{\dagger}}\gamma^5, 
\nonumber\\
\delta \bar{c}_{\psi}=-i\alpha \bar{c}_{\psi}\gamma^5, &&\quad
\delta\bar{c}_{\psi^{\dagger}}=i\alpha \gamma^5\bar{c}_{\psi^{\dagger}}. \nonumber 
\end{eqnarray}
The chiral transformations on $\psi$ and $\psi^{\dagger}$ are just the usual ones of Section 5, while the full set of Eq. (\ref{suptr}) can be considered the extension of the chiral transformations to the space of all the CPI fields. Using the standard trick of promoting $\alpha$ to be a local parameter, it is easy to prove
that the action $\widetilde{S}=\int dx \, \widetilde{\mathcal L}$, associated with the Lagrangian density (\ref{6-1}), changes according to the following equation:
\begin{equation}
\displaystyle \widetilde{S} \longrightarrow \; \widetilde{S} - i \int dx \left[
\alpha({{x}})\partial_{\mu}\left(\lambda_{\psi}\gamma^{0}
\gamma^{\mu}\gamma^{5}\psi-
\psi^{\dagger}\gamma^{0}\gamma^{5}
\gamma^{\mu}\lambda_{\psi^{\dagger}}+i\bar{c}_{\psi}\gamma^{0}
\gamma^{\mu}\gamma^{5}c^{\psi}
+ic^{\psi^{\dagger}}\gamma^{0}\gamma^{5}
\gamma^{\mu}\bar{c}_{\psi^{\dagger}}\right)\right]. \label{2.9}
\end{equation}
From (\ref{2.9}) we can read off the following current associated to the transformations (\ref{suptr}):
\begin{displaymath}
\widetilde{J}^{\mu}_5=-i\biggl[\lambda_{\psi}\gamma^{0}\gamma^{\mu}\gamma^{5}\psi-
\psi^{\dagger}\gamma^{0}\gamma^{5}
\gamma^{\mu}\lambda_{\psi^{\dagger}}+i\bar{c}_{\psi}\gamma^{0}\gamma^{\mu}\gamma^{5}c^{\psi}
+ic^{\psi^{\dagger}}\gamma^{0}\gamma^{5}\gamma^{\mu}\bar{c}_{\psi^{\dagger}}\biggr],
\end{displaymath}
while the associated charge is
\begin{equation}
\displaystyle \widetilde{Q}_5=\int d\mathbf{x}\biggl[-i\lambda_{\psi}(x)
\gamma^5\psi(x)-i\psi^{\dagger}(x)\gamma^5\lambda_{\psi^{\dagger}}(x)
+\bar{c}_{\psi}(x)\gamma^5c^{\psi}(x)
-c^{\psi^{\dagger}}(x)\gamma^5\bar{c}_{\psi^{\dagger}}(x)\biggr]. \label{supch}
\end{equation}
The chiral charge above is conserved, as we can prove explicitly by using the equations of motion. Such charge could have been obtained also by building, like in Eq. (\ref{supero}), the Lie derivative along the Hamiltonian vector field associated with the charge $Q_5$ of Eq. (\ref{qu5}). This object lifts, via the commutators defined in the CPI, the standard phase space chiral transformations to the extended ones given by Eq. (\ref{suptr}).

In Section 3 we have shown that we could build a different CPI formulation using as basic variables $\psi$ and $\bar{\psi}$ instead of $\psi$ and $\psi^{\dagger}$. Using the rules (\ref{rul}), it is easy to rewrite the chiral transformations (\ref{suptr}) as follows:
\begin{eqnarray}
\label{trapsi} \delta \psi=i\alpha \gamma^5\psi, &&\quad \delta \bar{\psi}=
i\alpha\bar{\psi}\gamma^5,\nonumber\\
\delta \lambda_{\psi}=-i\alpha \lambda_{\psi}\gamma^5, &&\quad \delta \lambda_{\bar{\psi}}
=-i\alpha \gamma^5 \lambda_{\bar{\psi}},\\
\delta c^{\psi}=i\alpha \gamma^5c^{\psi}, &&\quad \delta c^{\bar{\psi}}=i\alpha c^{\bar{\psi}}\gamma^5, 
\nonumber\\
\delta \bar{c}_{\psi}=-i\alpha \bar{c}_{\psi}\gamma^5, &&\quad
\delta\bar{c}_{\bar{\psi}}=-i\alpha \gamma^5\bar{c}_{\bar{\psi}}. \nonumber 
\end{eqnarray}
Under the local version of these transformations the action $\bar{S}=\int dx \; \bar{\mathcal L}$ associated with the Lagrangian density  (\ref{suplag2}), where the derivatives have been replaced by the covariant derivatives, changes as follows:
\begin{displaymath}
\displaystyle \bar{S}\; \longrightarrow \; \bar{S}-i\int dx \;
\alpha({{x}})\partial_{\mu}\left[\lambda_{\psi}\gamma^{0}\gamma^{\mu}\gamma^{5}\psi-
\bar{\psi}\gamma^{5}\gamma^{\mu}\gamma^{0}\lambda_{\bar{\psi}}+i\bar{c}_{\psi}\gamma^{0}
\gamma^{\mu}\gamma^{5}c^{\psi}+ic^{\bar{\psi}}\gamma^{5}\gamma^{\mu}
\gamma^{0}\bar{c}_{\bar{\psi}}\right].
\end{displaymath}
As before, one can immediately read off the associated conserved charge:
\begin{displaymath}
\displaystyle \bar{Q}_5=\int d\mathbf{x}\biggl[-i\lambda_{\psi}(x)
\gamma^5\psi(x)+i
\bar{\psi}(x)\gamma^5\lambda_{\bar{\psi}}(x)
+\bar{c}_{\psi}(x)\gamma^5c^{\psi}(x)
+c^{\bar{\psi}}(x)\gamma^5\bar{c}_{\bar{\psi}}(x)\biggr].
\end{displaymath}
This charge generates the chiral transformations (\ref{trapsi}) via the commutators (\ref{supcomm2}). 

The reader may be annoyed by the large number of fields appearing in the CPI. 
Nevertheless it is possible to prove, with a procedure similar to the one given in Ref. \cite{cinque}, that all the fields appearing in the CPI (\ref{final}) can be considered the components of the following superfield:
\begin{equation}
\Phi^{\psi^a}(x)\equiv\psi^a(x)+\theta c^a(x)-i\bar{\theta}\omega^{ab}\bar{c}_b(x)-\bar{\theta}\theta \omega^{ab}\lambda_b(x), \label{suppf}
\end{equation}
which depends not only on space-time $x$, but also on a couple of Grassmann variables $\theta,\bar{\theta}$. In terms of its two components $a=1,2$, Eq. (\ref{suppf}) can be written as:
\begin{equation}
\begin{array}{l}
\Phi^{\psi}(x,\theta,\bar{\theta})=\psi(x)+\theta c^{\psi}(x)-i\bar{\theta}\bar{c}_{\psi^{\dagger}}(x)
-\bar{\theta}\theta \lambda_{\psi^{\dagger}}(x) \medskip \\
\Phi^{\psi^{\dagger}}(x,\theta,\bar{\theta})=\psi^{\dagger}(x)+\theta c^{\psi^{\dagger}}(x)-
i\bar{\theta}\bar{c}_{\psi}(x)-\bar{\theta}\theta\lambda_{\psi}(x). \label{supf}
\end{array}
\end{equation}
Let us notice that the superfields of Eq. (\ref{supf}), differently from the ones appearing in Ref. \cite{quattro}-\cite{cinque}, are Grassmannian odd like the fields $\psi$ and $\psi^{\dagger}$ we started from. This means that, with respect to \cite{quattro}-\cite{cinque}, in all the proofs that we will perform we must take into account that 
$\psi$ and $\psi^{\dagger}$ have a different Grassmannian character with respect to the variables $\varphi^a$ of the point particle. 

With the definition (\ref{supf}) it is easy to prove that the Hamiltonian density $\widetilde{\cal H}$ of Eq. (\ref{supham}), appearing in the weight of the CPI (\ref{final}), can be derived from the usual Dirac Hamiltonian $H$ (\ref{hamdirac}) via the relation:
\begin{displaymath}
\displaystyle i\int d\theta d\bar{\theta} \, H(\Phi^{\psi},\Phi^{\psi^{\dagger}})=\int d\mathbf{x} \, \widetilde{\cal H}. \label{conn}
\end{displaymath} 
So, in order to get $\widetilde{\cal H}$, it is sufficient to replace in the Hamiltonian (\ref{hamdirac}) the fields with the superfields (\ref{supf}) and integrate over the Grasmmann variables $\theta$ and $\bar{\theta}$. The same procedure holds also for the symmetry charges. For example, if we replace, in the charge (\ref{qu5}), the fields with the superfields and we integrate over $\theta$ and $\bar{\theta}$, then we get just the charge of Eq. (\ref{supch}), which generates the chiral transformations in the CPI:
\begin{displaymath}
i\int d\theta d\bar{\theta}\, Q_5(\Phi^{\psi},\Phi^{\psi^{\dagger}})=\widetilde
{Q}_5. 
\end{displaymath}
Also the graded commutators of the CPI can be written in a compact form using the superfields (\ref{supf}). In fact Eq. (\ref{supcomm}) is completely equivalent to the following anticommutator among the superfields:
\begin{equation}
\Bigl[\Phi^{\psi_{\alpha}}(\mathbf{x},t,\theta,\bar{\theta}),\Phi^{\psi_{\beta}^{\dagger}}
(\mathbf{x}^{\prime},t,\theta^{\prime},\bar{\theta}^{\prime})\Bigr]=-i\delta_{\alpha\beta}
\delta(\mathbf{x}-\mathbf{x}^{\prime})\delta(\bar{\theta}-
\bar{\theta}^{\,\prime})\delta(\theta-\theta^{\prime}). \label{supfcom}
\end{equation}
Now it is also easy to understand the reason why $\widetilde{Q}_5$ is a conserved charge in the CPI. In fact we know that in quantum field theory, using the anticommutator 
$\displaystyle \Bigl[\psi_{\alpha}(\mathbf{x}),\psi^{\dagger}_{\beta}(\mathbf{x}^{\prime})\Bigr]
=\delta_{\alpha\beta}\delta(\mathbf{x}-\mathbf{x}^{\prime})$, we can prove that
\begin{displaymath}
\left[H(\psi,\psi^{\dagger}),Q_5(\psi,\psi^{\dagger}) \right]=0.
\end{displaymath}
Analogously in the CPI formalism from the anticommutator 
(\ref{supfcom}) we can derive that:
\begin{displaymath}
\displaystyle \left[ i\int d\theta d\bar{\theta}\, H(\Phi^{\psi},\Phi^{\psi^{\dagger}}),
i\int d\theta d\bar{\theta}\,Q_5(\Phi^{\psi},\Phi^{\psi^{\dagger}})\right]=0,
\end{displaymath}
which means that $\widetilde{Q}_5$ is a conserved charge.
The proof is identical to the quantum one, provided we enlarge the space-time $x$ to the ``superspace-time" $(x,\theta,\bar{\theta})$ and we
replace the fields $(\psi,\psi^{\dagger})$ with the superfields $(\Phi^{\psi},
\Phi^{\psi^{\dagger}})$. 
Also the chiral transformations of the CPI fields (\ref{suptr})
can be written in a compact form replacing in the usual 
transformations $\delta \psi=i\alpha \gamma^5 \psi$ and $\delta \psi^{\dagger}=
-i\alpha\psi^{\dagger}\gamma^5$ the fermion fields $\psi$ and $\psi^{\dagger}$ with the superfields (\ref{supf}):
\begin{equation}
\delta \Phi^{\psi}=i\alpha \gamma^{5}\Phi^{\psi}, \qquad \quad \delta \Phi^{\psi^{\dagger}}
=-i\alpha \Phi^{\psi^{\dagger}}\gamma^5. \label{chirrc}
\end{equation}
If we expand (\ref{chirrc}) in $\theta$, $\bar{\theta}$ we get exactly the set of transformations (\ref{suptr}). 

If we want to adopt the spinors $\psi$ and $\bar{\psi}$, instead of $\psi$ and $\psi^{\dagger}$, we can use Eq. (\ref{rul}) to replace the superfields (\ref{supf}) with the following ones:
\begin{displaymath}
\displaystyle \begin{array}{l}
\Phi^{\psi}(x, \theta,\bar{\theta})=\psi(x)+\theta c^{\psi}(x)-i\bar{\theta}\gamma^0\bar{c}_{\bar{\psi}}(x)-\bar{\theta}\theta
\gamma^0\lambda_{\bar{\psi}}(x) \medskip\\
\Phi^{\bar{\psi}}(x, \theta, \bar{\theta})=\bar{\psi}(x)+\theta c^{\bar{\psi}}(x)-i\bar{\theta}\bar{c}_{\psi}(x)\gamma^0-\bar{\theta}\theta \lambda_{\psi}(x)\gamma^0.
\end{array}
\end{displaymath}
Also in this case the chiral transformations (\ref{trapsi}) can be written in a compact form as:
\begin{displaymath}
\delta \Phi^{\psi}=i\alpha \gamma^5\Phi^{\psi}, \qquad \quad \delta
\Phi^{\bar{\psi}}=i\alpha\Phi^{\bar{\psi}}\gamma^5.  
\end{displaymath}

\section{Cancellation of the Chiral Anomaly in the CPI}

From Eq. (\ref{trapsi}) we note that the fields $(\psi,\bar{\psi})$ and $(c^{\psi},c^{\bar{\psi}})$
transform with opposite signs w.r.t. their conjugated momenta $(\lambda_{\psi},
\lambda_{\bar{\psi}})$ and $(\bar{c}_{\psi},\bar{c}_{\bar{\psi}})$. 
As we are going to prove now, this fact implies the cancellation of the chiral anomaly.
In fact, using the same arguments given by Fujikawa for the quantum path integral, we will prove that also in the CPI the functional measure in $\psi$ and $\bar{\psi}$
changes with the inverse of the determinant of the matrix $C$ written in Eq. (\ref{matrixc}). Nevertheless 
the functional measure for the associated conjugated momenta
$(\lambda_{\psi},\lambda_{\bar{\psi}})$ changes with the inverse of the determinant of
$C^{-1}$. Since the same happens in the two sectors of the variables $c$ and $\bar{c}$, we can say that this phenomenon will produce the cancellation of the chiral anomaly in the classical path integral.  

In order to prove in detail what we said above, let us consider that, according to the Hermiticity conditions (\ref{herm}),
$\lambda_{\psi}$ and $\lambda_{\bar{\psi}}$ are proportional to the Hermitian 
conjugate of $\psi$ and $\bar{\psi}$ respectively. Consequently if we write $\psi(x)$
as a linear combination of the orthonormal set of eigenfunctions $\{ \phi_n(x) \}$:
\begin{displaymath}
\displaystyle \psi(x)=\sum_nb_n\phi_n(x)
\end{displaymath}
then the field $\lambda_{\psi}(x)$ can be written as a linear combination of the fields 
$\{ \phi_n^{\dagger}(x) \}$:
\begin{displaymath}
\lambda_{\psi}(x)=\sum_n\phi_n^{\dagger}(x)\beta_n.
\end{displaymath}
As usual, we can derive the manner in which the coefficients $b_n$ and $\beta_n$ change under a chiral transformation directly from the transformations rules of the fields $\psi(x)$ and $\lambda_{\psi}(x)$. For example:
\begin{displaymath}
\begin{array}{l}
\displaystyle \psi^{\prime}(x)=\Bigl[1+i\alpha(x)\gamma^5\Bigr]\psi(x) \medskip \\
\displaystyle \Longrightarrow \, b_m^{\prime}=\sum_n\Bigl[\delta_{mn}+i\int dx \; \alpha(x) \phi_m^{\dagger}(x) \gamma^5\phi_n(x)\Bigr]b_n=\sum_nC_{mn}b_n,
\end{array}
\end{displaymath} 
where
\begin{equation}
C_{mn}\equiv \delta_{mn}+i\int dx \; \alpha(x)\phi_m^{\dagger}(x)\gamma^5\phi_n(x).
\label{7-0}
\end{equation}
The associated momenta $\lambda_{\psi}$ transform instead in the following manner:
\begin{displaymath}
\begin{array}{c}
\displaystyle \lambda_{\psi}^{\prime}(x)=\lambda_{\psi}(x)\Bigl[1-i\alpha(x)\gamma^5\Bigr] \medskip \\
\displaystyle \Longrightarrow\beta_m^{\prime}=
\sum_m\beta_m\Bigl[\delta_{mn}-i\int dx \; \alpha(x) \phi_m^{\dagger}(x)
\gamma^5\phi_n(x)\Bigr]=\sum_m\beta_mD_{mn},
\end{array}
\end{displaymath}
where $D_{mn}$ is the matrix:
\begin{equation}
\displaystyle D_{mn}\equiv \delta_{mn}-i\int d{x}\; \alpha({x}) \phi_m^{\dagger}({x})
\gamma^5\phi_n({x}). \label{matricedi}
\end{equation}
Using the same arguments, if we write the following expansions for the fields $\bar{\psi}$ and $\lambda_{\bar{\psi}}$:
\begin{displaymath}
\displaystyle \bar{\psi}(x)=\sum_m\phi_m^{\dagger}(x)\bar{b}_m,  \qquad\quad
\displaystyle \lambda_{\bar{\psi}}(x)=\sum_m\bar{\beta}_m\phi_m(x)
\end{displaymath}
we can derive from (\ref{trapsi})  that the coefficients $\bar{b}_m$
and $\bar{\beta}_m$ transform as follows:
\begin{displaymath}
\bar{b}_n^{\prime}=\sum_m\bar{b}_mC_{mn},\qquad \quad \bar{\beta}_m^{\prime}=\sum_nD_{mn}\bar{\beta}_{n}.
\end{displaymath}
Consequently the functional measure 
\begin{displaymath}
\displaystyle {\mathscr D}\psi{\mathscr D}\bar{\psi}{\mathscr D}\lambda_{\psi}
{\mathscr D} \lambda_{\bar{\psi}}=\prod_m db_md\bar{b}_md\beta_md\bar{\beta}_m
\end{displaymath}
changes under chiral transformations as follows:
\begin{displaymath}
\displaystyle {\mathscr D}\psi^{\prime}{\mathscr D}\bar{\psi}^{\prime}
{\mathscr D}\lambda_{\psi}^{\prime}{\mathscr
D}\lambda_{\bar{\psi}}^{\prime}=\widetilde{J} \,
\cdot {\mathscr D}\psi{\mathscr D}\bar{\psi}{\mathscr D}\lambda_{\psi}{\mathscr
D}\lambda_{\bar{\psi}}
\end{displaymath}
where $\widetilde{J}=[\textrm{det} \, C]^{-2}\,[\textrm{det} \, D]^{-2}$. Let us now notice that 
\begin{eqnarray}
\displaystyle (CD)_{nl}&=&\sum_m \Bigl(\delta_{nm}+i\int dx \; \alpha(x)
\phi_n^{\dagger}(x)\gamma^5\phi_m(x)\Bigr) \cdot\nonumber\\
&& \cdot \Bigl(\delta_{ml}-i\int dx \; \alpha(x) \phi_m^{\dagger}(x)\gamma^5\phi_l(x)\Bigr)=
\delta_{nl}+O(\alpha^2). \nonumber
\end{eqnarray}
This means that, except for terms of order $\alpha^2$, the matrix $D=C^{-1}$.
Therefore $[\textrm{det} \, D]^{-2}=[\textrm{det} \, C]^2$ and so the Jacobian of the transformation is equal to one: $\widetilde{J}=1$.
So we can conclude that the chiral transformation of the functional measure 
${\mathscr D}\psi{\mathscr D}\bar{\psi}$ is exactly compensated by the chiral transformation of ${\mathscr D}\lambda_{\psi}{\mathscr D}\lambda_{\bar{\psi}}$, i.e., the functional measure $\displaystyle {\mathscr D}\psi{\mathscr D}\bar{\psi}{\mathscr D}\lambda_{\psi}
{\mathscr D} \lambda_{\bar{\psi}}$ is invariant under chiral transformations. 

A similar argument holds also for the part of the measure involving the auxiliary variables $c$. 
In fact, if we expand the fields $c$ over an orthonormal and complete set of states:
\begin{displaymath}
\displaystyle  c^{\psi}({x})=\sum_na_n\varphi_n({x}),\qquad \quad
c^{\bar{\psi}}({x})=\sum_n\varphi_n^{\dagger}({x})\bar{a}_n 
\end{displaymath}
then the functional measure ${\mathscr D}c^{\psi}{\mathscr D}{c}^{\bar{\psi}}$ can be rewritten as a product of differentials of the expansion coefficients $\displaystyle \prod_nda_n d\bar{a}_n$. The infinitesimal chiral transformations (\ref{trapsi}) imply the following transformation equation:
\begin{displaymath}
\displaystyle \prod_nda_n^{\prime}d\bar{a}_n^{\prime}=[\textrm{det} \, A]^{2}
\prod_nda_nd\bar{a}_n, \qquad   A_{mn}\equiv \delta_{mn}+i\int
dx \; \alpha(x) \varphi_m^{\dagger}(x)\gamma^5\varphi_n(x).
\end{displaymath}
Consequently there appears a non-trivial Jacobian for the chiral transformation of the 
measure in the fields $c$:
\begin{displaymath}
\displaystyle {\mathscr D}c^{\psi^{\prime}}{\mathscr D}c^{\bar{\psi}^{\prime}}=[\textrm{det} \, A]^{2} \;
{\mathscr D}c^{\psi}{\mathscr D}c^{\bar{\psi}}.
\end{displaymath}
In this case there appears the square of a determinant instead of its inverse because the fields $c^{\psi}$ and $c^{\bar{\psi}}$ are Grassmannian even. Since also $\bar{c}_{\bar{\psi}}$ and $\bar{c}_{\psi}$ are scalar bosonic fields, like $c^{\psi}$ and $c^{\bar{\psi}}$, we can expand them on the same orthonormal basis $\{\varphi_n(x)\}$ and $\{\varphi_n^{\dagger}(x)\}$. What we finally get is:
\begin{displaymath}
\displaystyle {\mathscr D}\bar{c}_{\psi}^{\, \prime}{\mathscr D}\bar{c}_{\bar{\psi}}^{\, \prime}=
[\textrm{det} \, B]^{2}\,
{\mathscr D}\bar{c}_{\psi}{\mathscr D}\bar{c}_{\bar{\psi}},
\end{displaymath}
where the matrix $B$ is given by:
\begin{displaymath}
\displaystyle B_{mn}\equiv\delta_{mn}-i\int
dx \; \alpha(x)\varphi_m^{\dagger}(x)\gamma^5\varphi_n(x).
\end{displaymath}
The matrices $A$ and $B$ are one the inverse of the other, except for terms of order 
$\alpha^2$. Consequently $[\textrm{det} \,B]^2=[\textrm{det} \,A]^{-2}$ and the chiral transformation
of the functional measure ${\mathscr D}c^{\psi}{\mathscr D}c^{\bar{\psi}}$ is compensated by the chiral transformation of the functional measure 
${\mathscr D}\bar{c}_{\psi}{\mathscr D}\bar{c}_{\bar{\psi}}$.
So in
the CPI, differently than in the QPI, there is a ``cancellation" of the chiral anomaly just 
because of the presence of the auxiliary fields in the functional measure. 
Furthermore we can say that the cancellation of the anomaly takes place separately in each form-sector \cite{quattro}. In fact there is a cancellation of the anomaly in the zero-form sector, where only the variables $\psi, \bar{\psi},\lambda_{\psi}, \lambda_{\bar{\psi}}$ appear, while if we consider the variables $c$, $\bar{c}$ the cancellation takes place internally in the $c$, $\bar{c}$-sector, which represents the sector at higher form number. 

Up to now we have analyzed the behavior under chiral transformations of the functional measures of the quantum and classical path integrals for the spinors $\psi$ and $\bar{\psi}$.
What would have happened if instead we had considered the path integral in $\psi$ and $\psi^{\dagger}$? Let us start from the quantum case.
According to (\ref{suptr}) $\psi^{\dagger}$ changes as:
\begin{displaymath}
\displaystyle \psi^{\dagger^{\prime}}=\psi^{\dagger}(1-i\alpha\gamma^5).
\end{displaymath} 
So, if we expand the field $\psi^{\dagger}$ over the same basis $\{ \phi_n^{\dagger}\}$
on which we expanded $\bar{\psi}$:
\begin{displaymath}
\displaystyle \psi^{\dagger^{\prime}}(x)=\sum_n\phi_n^{\dagger}(x) \bar{\beta}_n^{\prime} 
\end{displaymath}
we would easily get the equation of transformation for the coefficients:
\begin{displaymath}
\displaystyle \bar{\beta}_m^{\prime}=
\sum_n \bar{\beta}_n \left[ \int dx \,\phi_n^{\dagger}(x)(1-i\alpha(x)\gamma^5)
\phi_m(x)\right]=\sum_n\bar{\beta}_n D_{nm},
\end{displaymath}
where $D$ is the matrix defined in Eq. (\ref{matricedi}).
Since the coefficients of the expansion of $\psi(x)$ transform with the matrix $C$ of Eq. (\ref{matrixc}) and $C$ and $D$ are one the inverse of the other, the whole change of the functional measure 
$\displaystyle {\mathscr D}\psi{\mathscr D}\psi^{\dagger}=\prod_ndb_nd\bar{\beta}_n$ is given by:
\begin{equation}
\displaystyle  \prod_ndb_n^{\prime}d\bar{\beta}_n^{\,\prime}=[\textrm{det} \, C]^{-1}\,[\textrm{det} \, D]^{-1}
\prod_ndb_nd\bar{\beta}_n= 
\prod_ndb_nd\bar{\beta}_n+O(\alpha^2). \label{6-2}
\end{equation}
This means that the Jacobian of the chiral transformation 
${\mathscr D}\psi {\mathscr D}\psi^{\dagger}\, \rightarrow {\mathscr D}\psi^{\prime} 
{\mathscr D}\psi^{{\dagger}^{\prime}}$ is equal to one, except for terms of order $\alpha^2$.
This is a consequence of the fact that, under the infinitesimal chiral transformations (\ref{suptr}), $\psi^{\dagger}$ transforms with the opposite
sign w.r.t. $\bar{\psi}$. So if we had used ${\mathscr D}\psi{\mathscr D}\psi^{\dagger}$ as {\it quantum} path integral measure, the result in (\ref{6-2}) would have implied that a cancellation of the chiral anomaly occurs also at the quantum level. Anyhow we must notice that the measure ${\mathscr D}\psi{\mathscr D}\psi^{\dagger}$ is not
acceptable as a {\it quantum} path integral measure because, differently from 
${\mathscr D}\psi{\mathscr D}\bar{\psi}$, it is not Lorentz covariant.
So we can say that at the quantum level it is possible to have a functional measure invariant under chiral transformations, provided we give up the Lorentz covariance. Vice versa a Lorentz covariant functional measure 
implies a chiral anomaly. 

The matrix $\gamma^{0}$
is crucial in order to go from the functional measure ${\mathscr D}\psi{\mathscr D}{\psi}^{\dagger}$
to the one given by ${\mathscr D}\psi{\mathscr D}\bar{\psi}$ and vice versa. In what follows we will show which is the proper manner to handle these matrices $\gamma^0$ in the path integral formalism. Let us consider the expansion of the two
fields $\bar{\psi}$ and $\psi^{\dagger}$ on the same set of eigenfunctions $\{ \phi^{\dagger}_m \}$:
\begin{displaymath}
\displaystyle \bar{\psi}=\sum_m\phi^{\dagger}_m\bar{b}_m,\qquad \qquad \psi^{\dagger}=\sum_m\phi_m^{\dagger}
\bar{\beta}_m. 
\end{displaymath}
The relationship between the coefficients $\bar{\beta}_m$ and $\bar{b}_m$ can be easily 
derived from the relationship between $\bar{\psi}$ and $\psi^{\dagger}$:
\begin{equation}
\displaystyle \bar{\psi}=\psi^{\dagger}\gamma^{0}
\; \Longrightarrow \; \sum_m\phi_m^{\dagger}\bar{b}_m=\sum_m\phi_m^{\dagger}\gamma^{0}\bar{\beta}_m. \label{comp2}
\end{equation}
Using (\ref{comp2}) we easily get:
\begin{displaymath}
\displaystyle \bar{b}_n=\sum_m\bar{\beta}_m \int d{x}\;\phi_m^{\dagger}({x})\gamma^{0} 
\phi_n({x})=\sum_m\bar{\beta}_m\Gamma_{mn}, \label{compp}
\end{displaymath}
where we have defined the following $\Gamma$ matrix:
\begin{equation}
\displaystyle \Gamma_{mn}\equiv\int d{x}\,\phi_m^{\dagger}({x})\gamma^{0}
\phi_n({x}). \label{Gammamat}
\end{equation}
This implies that ${\mathscr D}\bar{\psi}={\mathscr D}\psi^{\dagger}\,[\textrm{det}\,\Gamma]^{-1}$, i.e., in going from ${\mathscr D}{\psi}^{\dagger}$ to ${\mathscr D}\bar{\psi}$ 
there appears the inverse of the determinant of the matrix $\Gamma$. This quantity is not just the determinant of the $4 \times 4$ matrix $\gamma^0$ but it is an infinite dimensional determinant and it must be calculated via some regularization procedure like the Fujikawa determinant. In the case of the measure ${\mathscr D}\psi{\mathscr D}{\psi}^{\dagger}$ we had no anomaly at the quantum level while we had it with ${\mathscr D}\psi{\mathscr D}\bar{\psi}$, so it should be clear that it is just the $[\textrm{det} \,\Gamma]^{-1}$ that produces the chiral anomaly of the functional measure ${\mathscr D}\psi{\mathscr D}\bar{\psi}$.
To prove that let us notice that $\gamma^0$ and $\gamma^5$ anticommute, so we easily get that the matrices $C$, $D$ and $\Gamma$ of Eq. (\ref{7-0}), (\ref{matricedi}) and (\ref{Gammamat}) obey the following equation: $D\Gamma=\Gamma C$. Consequently from 
\begin{equation}
\displaystyle {\mathscr D}\bar{\psi}={\mathscr D}{\psi}^{\dagger}\, [\textrm{det} \,\Gamma]^{-1}
\qquad \quad {\mathscr D}\psi^{\dagger \prime}={\mathscr D}\psi^{\dagger}\, [\textrm{det} \,D]^{-1} \label{matgamma}
\end{equation}
we easily get that 
\begin{displaymath}
\displaystyle {\mathscr D}\bar{\psi}^{\prime}={\mathscr D}{\psi}^{\dagger}
\,[\textrm{det} \,(D\Gamma)]^{-1}={\mathscr D}\psi^{\dagger}\, [\textrm{det} \,(\Gamma C)]^{-1}
={\mathscr D}\bar{\psi} \,[\textrm{det} \,C]^{-1}.
\end{displaymath}
So while the functional measure ${\mathscr D}\psi^{\dagger}$ transforms with $[\textrm{det} \,D]^{-1}$, which cancels exactly the $[\textrm{det} \,C]^{-1}$ appearing in the transformation of the functional measure ${\mathscr D}\psi$, the presence of $\Gamma$ in Eq. (\ref{matgamma}) implies that also ${\mathscr D}\bar{\psi}$ transforms with $[\textrm{det} \,C]^{-1}$ and this is just what avoids the cancellation of the chiral anomaly at the quantum level.

Before concluding this section, let us go to the CPI for the fields $\psi$ and $\psi^{\dagger}$. First of all let us see what happens when we pass from ${\mathscr D}\bar{\psi}{\mathscr D}\lambda_{\bar{\psi}}$ to ${\mathscr D}\psi^{\dagger}{\mathscr D}\lambda_{\psi^{\dagger}}$.
We want to prove that also in this case no chiral anomaly arises. 
This means that the $\gamma^{0}$ matrices that allow us to go from  $(\bar{\psi},\lambda_{\bar{\psi}})$ to $(\psi^{\dagger},\lambda_{\psi^{\dagger}})$ according to the equations:
\begin{equation}
\bar{\psi}=\psi^{\dagger}\gamma^{0}, \qquad \quad \lambda_{\bar{\psi}}=
\gamma^{0}\lambda_{\psi^{\dagger}} \label{per}
\end{equation}
cannot generate any chiral anomaly in the CPI. 
In fact, while the equation of transformation between ${\mathscr D} \psi^{\dagger}$ and ${\mathscr D}\bar{\psi}$ is the one previously shown:
\begin{equation}
{\mathscr D}\bar{\psi}={\mathscr D}\psi^{\dagger}\;[\textrm{det} \,\Gamma]^{-1},
\label{gammauno}
\end{equation}
for the fields $\lambda_{\psi^{\dagger}}$, $\lambda_{\bar{\psi}}$ we have:
\begin{displaymath}
\lambda_{\bar{\psi}}=\gamma^{0}\lambda_{\psi^{\dagger}}\;\Longrightarrow \;
\sum_n a_n\phi_n=\sum_n\gamma^{0}\alpha_n\phi_n,
\end{displaymath} 
which implies the following equation for the coefficients:
\begin{displaymath}
\displaystyle a_m=\sum_n \Gamma_{mn}\alpha_n.
\end{displaymath}
This means that also in going from ${\mathscr D}\lambda_{\psi^{\dagger}}$ to ${\mathscr D}\lambda_{\bar{\psi}}$ there appears the inverse of the determinant of the matrix $\Gamma$ of Eq. (\ref{Gammamat}):
\begin{equation}
{\mathscr D}\lambda_{\bar{\psi}}=[\textrm{det} \,\Gamma]^{-1} \, {\mathscr D}\lambda_{\psi^{\dagger}}. \label{gammadue}
\end{equation}
Collecting together (\ref{gammauno}) and (\ref{gammadue}) we have that: 
\begin{equation}
{\mathscr D}\bar{\psi}{\mathscr D}\lambda_{\bar{\psi}}=
{\mathscr D}\psi^{\dagger}\, [\textrm{det} \,\Gamma^2]^{-1}\,{\mathscr D}\lambda_{\psi^{\dagger}}. \label{trasf}
\end{equation}
Using the fact that $\gamma^0$ is idempotent it is easy to prove that also $\Gamma^2=\mathbb{I}$. In fact:
\begin{eqnarray}
\displaystyle \Gamma^2_{nl}&=&\sum_m\int d{x}\,\phi_n^{\dagger}({x})\gamma^{0}
\phi_m(x)\int d{y}\,\phi_m^{\dagger}({y})
\gamma^{0}\phi_{l}({y})=\nonumber\\
&=&\int d{x}d{y} \,\phi_n^{\dagger}({x})\gamma^{0}
\delta({x}-{y})\gamma^{0}\phi_l(y)= \nonumber \\
&=&\int d{x} \,\phi_n^{\dagger}(x)(\gamma^{0})^2\phi_l(x)
=\int d{x} \,\phi_n^{\dagger}({x})\phi_{l}({x})=\delta_{nl}. \nonumber
\end{eqnarray}
So we get that: $[\textrm{det} \, \Gamma^2]^{-1}=[\textrm{det}\,\mathbb{I}]^{-1}=1$ which, inserted in (\ref{trasf}), implies that we can go from ${\mathscr D}\bar{\psi}{\mathscr D}\lambda_{\bar{\psi}}$ to ${\mathscr D}\psi^{\dagger}{\mathscr D}\lambda_{\psi^{\dagger}}$ without generating any extra term.
Since the equations of transformation among the variables $c$ and $\bar{c}$ have the same form as the ones of (\ref{per}):
\begin{displaymath}
c^{\bar{\psi}}=c^{\psi^{\dagger}}\gamma^{0}, \qquad \bar{c}_{\bar{\psi}}=
\gamma^{0}\bar{c}_{\psi^{\dagger}}
\end{displaymath}
we can immediately say that also the Jacobian of the transformation:
\begin{displaymath}
{\mathscr D}c^{\bar{\psi}}{\mathscr D}\bar{c}_{\bar{\psi}}\,\longrightarrow \, {\mathscr D}c^{\psi^{\dagger}} {\mathscr D}\bar{c}_{\psi^{\dagger}}
\end{displaymath}
is equal to one. 

So, after having shown that the functional measure of the CPI (\ref{pathz}) for the fields $\psi,\bar{\psi}$ is invariant under chiral transformations, we have proved the invariance also of the functional measure of the CPI (\ref{final}) for the fields $\psi, \psi^{\dagger}$,
using the properties of the matrix $\Gamma$. Of course we could have proven the invariance of the functional measure also by applying directly the Fujikawa's method. The result would have been the same, as the reader can easily realize. We can then conclude that no chiral anomaly arises at the classical level, irrespectively of which classical path integral we use. 

\section{Conclusions}

The superficial reader may consider this paper just an ``exercise" on the
Fuijkawa approach to anomalies. We do not think so for three different
reasons.

The first one goes back to the original work of Fujikawa which, when it first appeared,
raised several questions in the readers. The most common question was ``{\it The measure ${\mathscr D}\bar{\psi}{\mathscr D}\psi$ is a purely geometrical object, why should it contain any quantum information?}" and
an answer was ``{\it It gets  the quantum nature via the expansion  in the modes of the quantum Dirac operator}". We do not think this was the right answer because even in our classical case we could
expand the measure in the modes of the Dirac operator without getting any anomaly. We feel that the right answer to this question comes from the present paper and it is the following: ``{\it The Fujikawa measure ${\mathscr D}\bar{\psi}{\mathscr D}\psi$ was  already  the quantum measure because the classical one is not ${\mathscr D}\bar{\psi}{\mathscr D}\psi$ but the one we used in this paper ${\mathscr D}\bar{\psi} {\mathscr D}\psi {\mathscr D}\lambda_{\bar{\psi}} {\mathscr D}\lambda_{\psi}
{\mathscr D}c^{\bar{\psi}}{\mathscr D}c^{\psi}{\mathscr D}\bar{c}_{\bar{\psi}}{\mathscr D}\bar{c}_{\psi}$}". They are these extra fields $\lambda$, $c$, $\bar{c}$ which ``cancel" the quantum effect.

The second reason for this paper is that it was crucial to test that, even in our different approach to classical physics, no chiral anomaly appears. If that had not been the case, this would have thrown seriuos doubts on the equivalence between the CPI and the standard formulation of CM.

The third reason is that the mechanism of cancellation of anomalies is a crucial ingredient in many sectors of theoretical physics ranging from the standard model to strings. The case analyzed in this paper is a new  example of cancellation of anomalies which has never been studied before in the literature.

We feel moreover that in our case this mechanism has a geometrical background because the set of auxiliary fields
$(\lambda,c,\bar{c})$, needed for the cancellation of the anomaly, makes up a structure known as a double bundle over phase space which is described in detail in Ref. \cite{regini}. For the reader familiar with Refs. \cite{quattro}-\cite{regini}, it will be immediately clear that the cancellation is due to the geometrical nature of these auxiliary variables. 

What we would like to do next is to repeat this analysis for the scale anomaly. We feel that this exercise may reveal not only a nice mathematical background but may also lead to some {\it physical} developments. Work on this project is in progress.

\section*{Acknowledgments}

We acknowledge helpful discussions with M. Reuter. This work has been supported by grants from MIUR, INFN and University of Trieste. A.S. acknowledges a teaching assistantship from Syracuse University.

\bigskip 
\bigskip 

\begin{center}
{\LARGE\bf Appendices}
\end{center}

\appendix
\makeatletter
\@addtoreset{equation}{section}
\makeatother
\renewcommand{\theequation}{\thesection.\arabic{equation}}

\section{Appendix }

In this Appendix we will construct, starting from $\lambda_{\psi}$ and $\lambda_{\bar{\psi}}$, two fields $\lambda$ and $\bar{\lambda}$ transforming respectively via $S$ and $S^{-1}$ under a Lorentz transformation. Furthermore, we will build the Lorentz invariant CPI in terms of these new variables. 

Remembering that $\lambda_{\psi}$ and $\lambda_{\bar{\psi}}$ transform according to Eq. (\ref{4-4}), if we define  
$\bar{\lambda}\equiv\lambda_{\psi}\gamma^{0}$ and $\lambda\equiv\gamma^{0}\lambda_{\bar{\psi}}$ their Lorentz transformations are the following ones:
\begin{equation}
\left\{
\begin{array}{l}
\bar{\lambda}^{\prime}=\lambda_{\psi}^{\prime}\gamma^{0}=\lambda_{\psi}S^{\dagger}
\gamma^{0}=\lambda_{\psi}\gamma^{0}
S^{-1}=\bar{\lambda}S^{-1} \medskip \\ 
\lambda^{\prime}=\gamma^{0}\lambda_{\bar{\psi}}^{\prime}=\gamma^{0} 
(S^{-1})^{\dagger}\lambda_{\bar{\psi}}=S\gamma^{0}\lambda_{\bar{\psi}}=S\lambda, \label{24}
\end{array}
\right.
\end{equation}
where we have used the property of the matrix $S$ given by Eq. (\ref{propesse}). Using $\lambda$ and $\bar{\lambda}$ the Grassmann part of the Lagrangian density can be rewritten as:
\begin{displaymath}
\bar{\cal L}_g=\bar{\lambda}\gamma^{\mu}\partial_{\mu}\psi-(\partial_{\mu}\bar{\psi})\gamma^{\mu}\lambda.
\end{displaymath}
Because of the transformations (\ref{24}) and the fact that $S^{-1}\gamma^{\mu}S=a^{\mu}_{\;\nu}\gamma^{\nu}$
this Lagrangian is manifestly invariant under the Lorentz transformations. 
In terms of the new variables $\lambda$ and $\bar{\lambda}$ the Grassmann part of the CPI can be written as:
\begin{displaymath}
\displaystyle Z_{\scriptscriptstyle \textrm{CM}}^{\prime}[0]=\int{\mathscr D}\psi{\mathscr D}\bar{\psi}{\mathscr D}
\lambda{\mathscr D}\bar{\lambda} \;\textrm{exp}\,\biggl[
i\int\, dx\,\biggl(\bar{\lambda}\gamma^{\mu}\partial_{\mu}\psi-
(\partial_{\mu}\bar{\psi})\gamma^{\mu}\lambda\biggr)
\biggr].
\end{displaymath}
This path integral can be considered as the result of the exponentiation of the Lorentz covariant equations of motion $\gamma^{\mu}\partial_{\mu}\psi=0$ and $(\partial_{\mu}\bar{\psi})\gamma^{\mu}=0$ via the Grassmann variables $\bar{\lambda}$ and $\lambda$ respectively. The reader may wonder why there is no functional Jacobian in the new measure ${\mathscr D}\lambda{\mathscr D}\bar{\lambda}$ with respect to the old one ${\mathscr D}\lambda_{\psi}{\mathscr D}\lambda_{\bar{\psi}}$. The reason is that the transformation equations $\bar{\lambda}=\lambda_{\psi}\gamma^0$ and $\lambda=\gamma^0\lambda_{\bar{\psi}}$ have the same form as the ones of Eq. (\ref{per}). Like in that case, the change of variables in the functional measure will generate a $[\textrm{det} \,\Gamma^2]^{-1}$. As we proved in Section 7, $\Gamma^2={\mathbb{I}}$ and so $[\textrm{det}\, \Gamma^2]^{-1}=1$. We can thus conclude that no extra term arises in going from ${\mathscr D}\lambda_{\psi}{\mathscr D}\lambda_{\bar{\psi}}$ to ${\mathscr D}\lambda {\mathscr D}\bar{\lambda}$. 

\section{Appendix}

In this Appendix we will prove that also the CPI for the fields $\psi$ and $\psi^{\dagger}$ is invariant under Lorentz transformations. 

First of all let us consider that, from the Lorentz transformations of $\bar{\psi}$ and $\lambda_{\bar{\psi}}$, i.e., 
\begin{displaymath}
\bar{\psi}^{\prime}(x^{\prime})=\bar{\psi}(x)S^{-1}, \qquad \quad \lambda_{\bar{\psi}}^{\prime}(x^{\prime})=(S^{\dagger})^{-1}\lambda_{\bar{\psi}}(x),
\end{displaymath}
we can easily deduce, via Eq. (\ref{rul}), the way $\psi^{\dagger}$ and $\lambda_{\psi^{\dagger}}$ transform:
\begin{equation}
\left\{
\begin{array}{l}
\psi^{\dagger \prime}(x^{\prime})=\bar{\psi}^{\prime}(x^{\prime})\gamma^0=\bar{\psi}(x)S^{-1}\gamma^0=\bar{\psi}(x)\gamma^0S^{\dagger}=\psi^{\dagger}(x)S^{\dagger}, \medskip \\
\lambda_{\psi^{\dagger}}^{\prime}(x^{\prime})=\gamma^0\lambda_{\bar{\psi}}^{\prime}(x^{\prime})=\gamma^0(S^{\dagger})^{-1}\lambda_{\bar{\psi}}(x)=S\gamma^0\lambda_{\bar{\psi}}(x)=S\lambda_{\psi^{\dagger}}(x).
\end{array}
\right.
\label{lortr}
\end{equation}
The fields $\psi$ and $\lambda_{\psi}$ transform instead in the usual way:
\begin{equation}
\psi^{\prime}(x^{\prime})=S\psi(x), \qquad \quad \lambda_{\psi}^{\prime}(x^{\prime})=\lambda_{\psi}(x)S^{\dagger}. 
\label{lortr2}
\end{equation}
By comparing (\ref{lortr}) with (\ref{lortr2}) we notice that under the Lorentz transformations $\lambda_{\psi}$ and $\lambda_{\psi^{\dagger}}$ transform respectively like $\psi^{\dagger}$ and $\psi$. So the functional measure ${\mathscr D}\psi{\mathscr D}\psi^{\dagger}{\mathscr D}\lambda_{\psi}{\mathscr D}\lambda_{\psi^{\dagger}}$ does not seem to be Lorentz invariant. Nevertheless we must take into account that in the CPI (\ref{final}) there is also a part of the functional measure that depends on the auxiliary fields $c$ and $\bar{c}$. From (\ref{ciuno}) and (\ref{cidue}) we get that $c^{\psi}$ and $\bar{c}_{\psi}$ transform just as $\psi$ and $\lambda_{\psi}$:
\begin{displaymath}
c^{\psi\prime}(x^{\prime})=S c^{\psi}(x), \qquad \quad \bar{c}_{\psi}^{\,\prime}(x^{\prime})=\bar{c}_{\psi}(x)S^{\dagger},
\end{displaymath}
while the equations of transformation for $\bar{c}_{\psi^{\dagger}}$ and $c^{\psi^{\dagger}}$, which can be deduced using Eq. (\ref{rul}), are just the same as the ones for $\lambda_{\psi^{\dagger}}$ and $\psi^{\dagger}$:
\begin{displaymath}
\bar{c}_{\psi^{\dagger}}^{\,\prime}(x^{\prime})=S\bar{c}_{\psi^{\dagger}}(x), \qquad \quad c^{\psi^{\dagger \prime}}(x^{\prime})=c^{\psi}(x)S^{\dagger}.
\end{displaymath}
The crucial observation to make at this point is that the fields $(\psi,\psi^{\dagger},\lambda_{\psi},\lambda_{\psi^{\dagger}})$ are Grassmannian odd, while the associated fields $(c^{\psi},c^{\psi^{\dagger}},\bar{c}_{\psi},\bar{c}_{\psi^{\dagger}})$, which obey just the same Lorentz transformations, are Grassmannian even. This immediately implies that, at the level of functional measure, while the Lorentz transformation from ${\mathscr D}\psi$ to ${\mathscr D}\psi^{\prime}$ generates the {\it inverse} of the functional determinant of the matrix $S$, the Lorentz transformation from ${\mathscr D}c^{\psi}$ to ${\mathscr D}c^{\psi\prime}$ generates instead the functional determinant of the same matrix $S$. So, provided the functional determinants are evaluated using the same regularization procedure, they cancel against each other. Since the same happens for all the other couples of Grassmannian odd/even fields, we can say that also the functional measure of the CPI in $\psi$, $\psi^{\dagger}$ is invariant under the Lorentz transformations differently than at the quantum level. 

The Lorentz invariance of the Lagrangian $\widetilde{\cal L}$ of Eq. (\ref{lagcomp}), which appears in the weight of the CPI for $\psi$ and $\psi^{\dagger}$, is guaranteed by the Lorentz invariance of $\bar{\cal L}$ which is just equal to $\widetilde{\cal L}$, once we perform the change of variables (\ref{rul}). So also the CPI in (\ref{final}) is Lorentz covariant and it is meaningful to consider it in our discussion on the chiral symmetry at the classical level.

\end{document}